\documentclass[seceq]{ptptex}

\usepackage{graphicx}
\usepackage{bm}
\usepackage{multirow}

\newcommand{\Slash}[1]{\ooalign{\hfil/\hfil\crcr$#1$}}
\newcommand{\eqn}[1]{\label{eq:#1}}
\newcommand{\refeq}[1]{(\ref{eq:#1})}

\newcommand{\Eq}{Eq.~\refeq}
\newcommand{\Eqs}[2]{Eqs.~(\ref{eq:#1}) and (\ref{eq:#2})}
\newcommand{\Eqss}[2]{Eqs.~(\ref{eq:#1})--(\ref{eq:#2})}
\newcommand{\vecbox}[1]{\mbox{\boldmath{$#1$}}}

\def\vlowk{$V_{low\, k}$}

\def\piless{EFT($\Slash{\pi}$)}
\def\pislash{ {\pi\hskip-0.6em /} }

\def\pislashx{ {\pi\hskip-0.45em /} }
\def\nopiy{EFT$(\pislash)$}
\def\nopix{ EFT(\pislashx) }
\def\veft{$V_{EFT}$}
\def\vph{$V_{ph}$}
\def\vpiI{$V_{EFT(\pi)}^{I}$}
\def\vpilessI{$V_{\nopix}^{I}$}
\def\vpiII{$V_{EFT(\pi)}^{II}$}
\def\vpilessII{$V_{\nopix}^{II}$}
\def\vpi{$V_{EFT(\pi)}$}
\def\vpiless{$V_{\nopix}$}

\def\pieft{EFT($\pi$)}

\def\eftpi{EFT($\pi$)}

\def\1s0{{}^1S_0}
\def\ts1{{}^3S_1}
\def\td1{{}^3D_1}
\def\ltap{\ \raise.3ex\hbox{$<$\kern-.75em\lower1ex\hbox{$\sim$}}\ }
\def\gtap{\ \raise.3ex\hbox{$>$\kern-.75em\lower1ex\hbox{$\sim$}}\ }
\def\ket#1{\vert#1\rangle}
\def\bra#1{\langle#1\vert}

\notypesetlogo                       

\title{
Consideration of Nuclear Forces\\
from the Viewpoint of the Renormalization Group
}

\subtitle{Relation of Nuclear Forces between Phenomenological Approach\\
and Nuclear Effective Field Theory}    

\author{
Satoshi X. \textsc{Nakamura}\footnote{E-mail:
nakkan@rcnp.osaka-u.ac.jp}%
}

\inst{
Theory Division,
Research Center for Nuclear Physics,
Osaka University\\
Ibaraki, 567-0047, Japan
}

\recdate{November 30, 2004}

\abst{
A relation between nuclear forces derived using a phenomenological
approach and nuclear
effective field theory (NEFT) is proposed from a renormalization group
point of view.
A phenomenological nuclear force (\vph) and
an NEFT-based $NN$-potential (\veft)
are satisfactorily related to each other through the Wilsonian
renormalization group (WRG) method. 
It is clearly shown that
use of the simple contact interactions in NEFT 
is adequate to simulate small scale phenomena, and that
an NEFT-based $NN$-potential (\veft) is free from
dependence on the model used to describe small scale phenomena.
We study the characteristics of \veft\ from a WRG point of view, emphasizing
points that have not previously been fully recognized.
We also use the \vlowk\ method and a unitary transformation method to
relate \vph\ to \veft.
It is found that they are not appropriate for this purpose.
}

\begin{document}

\maketitle

\section{Introduction}\label{sec_intro}
The nature of the nuclear force is one of the oldest subjects in nuclear physics.
In one widely used approach to the nuclear force,
one employs the one-pion-exchange potential
(OPEP) as the well-known long range mechanism 
and uses a phenomenological model to describe the
short-distance mechanism, which is not well known.
We refer to this type of nuclear force as a phenomenological
nuclear force.
At present, there are some high-precision phenomenological
$NN$-potentials, such as the CD-Bonn potential\cite{bonn} and the
Nijmegen potential.\cite{nij}

A different approach to the nuclear force, based on
nuclear effective field theory (NEFT), 
was proposed in Weinberg's seminal work\cite{weinberg} and has been
studied extensively.\footnote{
We consider NEFT in the form proposed by
Weinberg\cite{weinberg} in this work.
In another NEFT,\cite{pds} the
nuclear potential is not explicitly derived.
We do not discuss this type of NEFT in this work.
}
The NEFT approach to the nuclear force is described in detail in
several references.\cite{kolck_review}
NEFT has attracted much
interest because \veft
\footnote{
Hereafter, we denote an NEFT-based potential by \veft.
Similarly, we denote a phenomenological potential by \vph.
}
has the following desirable formal features that are not possessed by
\vph.
The first feature, which represents
the most important aspect of this approach, is that it uses a chiral effective
Lagrangian and therefore describes a nuclear
force in a manner consistent with the symmetry properties of QCD,
in particular the spontaneously broken chiral symmetry.
The Lagrangian is composed of effective degrees of freedom (d.o.f.) of
hadrons and is the most general,
as long as the assumed set of symmetries exists.
When we consider low-energy $NN$-scattering, the nucleon and the pion
are the effective d.o.f.; heavier d.o.f. (heavy mesons, $\Delta$, etc.)
are integrated out.
The second feature is that
the construction of the nuclear force from the Lagrangian
is systematic, following a power counting
rule.
With these two features,
NEFT is considered to be model independent.
Regarding phenomenology,
several authors have constructed \veft's and have shown their usefulness
in reproducing low-energy $NN$-data.\cite{kolck-ptl,epelbaum,idaho,NN-park}

Despite the desirable characteristics of NEFT,
some naive but yet unanswered questions may be raised.
One question concerns the use of a contact interaction.
In NEFT, it is believed that the effect of d.o.f. that are integrated out is
absorbed by the contact interactions between hadrons representing the effective d.o.f.
However, the structure of the contact interaction seems to be too simple to
simulate physics of small scales, when 
compared to those used in
phenomenological models. Is the use of the contact interaction sufficient
to simulate small scale physics?
The other question concerns the model-independence of NEFT.
Why is \veft\ so special among the many $NN$-potentials?
Is it still one of the many phase-equivalent potentials?
In fact, the claim that NEFT is model independent is based only on a qualitative
argument and no quantitative analysis regarding this matter has been made.
It is very desirable to change this situation.

The key to answering the above questions is to note the size of the model space
on which an $NN$-potential acts.
In constructing \veft, one uses a cutoff function to restrict the momentum
states which the nucleon occupies; \veft\ acts on a model space.\footnote{
We refer to a model space with an infinite cutoff as the ``full space''.
}
A typical size of the model space for \veft\ is considerably smaller than
that for \vph;
the model space for \vph\ is typically
rather large ($\sim$ a few GeV) compared to the energy region of interest
($\sim$ a few hundred MeV).
From a renormalization group (RG) point of view, the use of a smaller
model space implies a rougher description of the system; that is,
the system is described by a theory which includes less
{\it detail} regarding small scale physics.
Thus we can formulate a scenario for the relation between \veft\ and \vph\ as follows.
There are a number of the \vph's which reproduce the low-energy $NN$-data.
They differ from one another with regard to the ways in which they model
small scale physics.
We start with such  \vph's and integrate out the nucleon high-momentum
states, thereby reducing the size of their model spaces.
As the model space becomes smaller, the corresponding potential comes
to have less information about
details of the small scale physics. 
Therefore, it is expected that all \vph's eventually converge to a single
$NN$-potential with a sufficiently reduced model space.
In this way, the model dependence of \vph\ arising from the
treatment of small scale physics disappears.
Furthermore, the small scale part of that single potential is
expected to be accurately taken into account by simple contact interactions,
because the {\it details} of the small scale physics are no longer
important.
After all, \veft\ constitutes a
parameterization of the single $NN$-potential.
The usefulness of NEFT may be assessed by examining how well
the NEFT-based parametrization of the single $NN$-potential works.
The single model-space potential is free from
dependence on the details of the modeling of small scale
physics, and therefore
its parameterization, \veft, is also model independent in this sense.
This is our scenario for the relation between \veft\ and \vph.
If this scenario is shown to be valid, 
the following consequences may be realized.
First, we can answer the questions posed in the previous
paragraph and obtain a deeper understanding of NEFT.
Second, understanding the relation between \vph\ and \veft, we can
evaluate the role of NEFT in nuclear physics from a new point of view.
These expected consequences provide a good motivation for studying this
scenario.

The purpose of this work is to show that the above-described
scenario is indeed valid
and thereby to propose a relation between \veft\ and \vph.
For this purpose, we demonstrate the procedure presented above;
we start with some \vph's and perform a model-space reduction. Then we
examine whether the obtained potential is free from the model dependence
of the \vph\ and if it is accurately simulated by the NEFT-based parameterization,
\veft. The remaining problem
in the demonstration is to determine how we reduce the model space.
Some model-space reduction schemes have been proposed to this time;
the Wilsonian renormalization group method,\cite{wilson,birse}
the \vlowk\ method,\cite{vlowk} and a unitary transformation
method.\cite{UT}
However, it has not yet been established which should be used in
NEFT.
We will try each of these methods in our demonstration and find the
proper one.
The criterion for finding the proper method is
consistency with the basic ideas of NEFT.
We examine this consistency with respect to two points.
First,
the proper method should be consistent with the method for integrating out
the d.o.f. in NEFT; an effective Lagrangian is obtained
by integrating out the heavier d.o.f. using the path integral method.
As a second check of consistency,
the obtained potential with the reduced model space should be accurately
simulated by the NEFT-based parameterization.
In other words, 
the obtained potential should exhibit behavior such that
the low-energy constants (LECs: coupling
constants involved in an effective Lagrangian for NEFT) 
are {\it natural} and
the NEFT-based perturbation 
is {\it systematic}.
{\it Natural} LECs and a {\it systematic} perturbation
(where the meanings of ``natural'' and ``systematic''
are specified below)
are an important basis of NEFT for a
convergent perturbation scheme.
In fact, this criterion can be regarded as a test for examining whether 
NEFT itself is {\it natural} and {\it systematic}.
This is a new type of test for examining the fundamental aspects of NEFT.
We discuss this point.

We now describe the organization of the following sections.
In \S \ref{sec_rg},
we describe the possible methods for the model-space reduction scheme. 
We also discuss the properties of
the model-space potentials provided by these methods.
In \S \ref{sec_natural}, we discuss 
fundamental aspects of NEFT, such as its
{\it naturalness}.
In \S \ref{sec_veft}, we give explicit expressions for \veft\ to be
used in this work.
In \S \ref{sec_result}, we present model-space potentials obtained from
\vph's using the three methods.
Comparing these, we
find the proper model-space reduction scheme.
In \S \ref{sec_discussion},
we discuss the characteristics of \veft,
emphasizing points that have not yet been fully
recognized.
Finally, we give a conclusion in \S \ref{sec_conclusion}.

\section{Formalism}\label{sec_form}
\subsection{Model-space reduction}\label{sec_rg}

In this section, we will describe the possible methods for the model-space
reduction.
First, we attempt to find a reduction scheme
by considering the construction of an effective Lagrangian.
We start with a Lagrangian ${\cal L}_H$ in which some d.o.f. $\Psi$ appear
explicitly. (The subscript ``$H$'' means ``heavy''.)
The S-matrix element for a given process is obtained from the path
integral $Z$:
\begin{eqnarray}
 \eqn{z_lh}
 Z &=& \int {\cal D}\Psi e^{i\int  d^4\!x\ {\cal L}_H}\ .
\end{eqnarray}
Now we divide the d.o.f. into ``heavy'' ($\Psi_H$) and ``light''
($\Psi_L$) d.o.f. 
and suppose that we are interested in a system with an energy
scale for which only ``light'' d.o.f. are important.
In this case, the idea of EFT is to obtain an
effective Lagrangian written in terms of only ``light'' d.o.f.
Therefore, we integrate out the ``heavy'' d.o.f., obtaining
\begin{eqnarray}
\eqn{z_ll}
 Z &=& \int {\cal D}\Psi_L e^{i\int  d^4\!x\ {\cal L}_L}\ ,
\end{eqnarray}
where ${\cal L}_L$ ($L$ means ``light'') is the effective Lagrangian.

\begin{figure}[t]
\begin{center}
 \includegraphics[width=80mm]{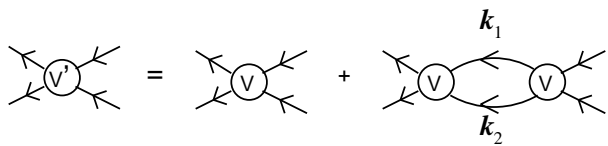}
\end{center}
\caption{\label{fig_rge} 
For an infinitesimal reduction of the cutoff,
the WRG equation represents the evolution of the $NN$-interaction, as
graphically shown
above. $V$ ($V'$) is the original (renormalized) interaction. In the
loop, the nucleonic momenta are denoted by $\vecbox{k}_1$ and
$\vecbox{k}_2$, which lie in the momentum shell that is integrated out.
}
\end{figure}

The reduction of the model space for the nucleon states also
should be carried out following the path integral discussed above.
This is the Wilsonian renormalization group (WRG) method.\cite{wilson}
In general, we cannot fully carry out the integration, 
because there are an infinite number of and various types of terms in the
Lagrangian.
Even so, 
as long as we are concerned with low-energy $NN$-scattering and
the $NN$-interaction is provided by an $NN$-potential,
we can manage to perform the integration.
In this case, we can start with the path integral given in \Eq{z_lh},
where $\Psi$ is the nucleon field that includes the nucleon momentum
states restricted by a cutoff, $\Lambda_H$.
The Lagrangian ${\cal L}_H$ includes the starting $NN$-potential, $V_H$.
We separate the nucleonic d.o.f. into high-momentum ($\Psi_H$) and
low-momentum states ($\Psi_L$) by a cutoff $\Lambda_L$.
Then we integrate out the high-momentum states to arrive at \Eq{z_ll},
in which the effective Lagrangian ${\cal L}_L$ includes the low-momentum
$NN$-potential, $V_L$.
In fact, if we work in the center of mass (CM) $NN$-system,
the evolution of the $NN$-potential driven by the integration is given
by the WRG equation (See Appendix \ref{app_rg} for a derivation.)
\begin{eqnarray}
\eqn{rge}
 {\partial V^{(\alpha)}(k',k;p,\Lambda) \over\partial\Lambda} 
= {M\over 2\pi^2} V^{(\alpha)}(k',\Lambda;p,\Lambda){\Lambda^2\over\Lambda^2-p^2}
V^{(\alpha)}(\Lambda,k;p,\Lambda)\ ,
\end{eqnarray}
where $V^{(\alpha)}$ is the $NN$-potential for a given channel (partial wave)
$\alpha$, and $M$ denotes the nucleon mass.
In $V^{(\alpha)}(k',k;p,\Lambda)$,
$\Lambda$ is a cutoff for the one-nucleon momentum,
$p$ is an on-shell one-nucleon momentum, $p\equiv\sqrt{ME}$ with $E$ being
the kinetic energy of the two nucleons,
and $k$ ($k'$) is a one-nucleon momentum before (after) the interaction.
Note that we are working in the CM system, and therefore the magnitude of
the one-nucleon momentum is the same as that of the relative momentum of the
two-nucleon system.
We use the same notation throughout this work.
The WRG equation given by \Eq{rge} is for a single channel $\alpha$, but the
extension to the coupled-channel case is straightforward.
Equation \refeq{rge} is
graphically shown in Fig.~\ref{fig_rge}.
For an infinitesimal reduction of the cutoff, the interaction $V$
evolves into the renormalized one $V'$
by absorbing the one-loop graph.
In the figure, the loop diagram includes
the intermediate one-nucleon states 
of $\Lambda-\delta\Lambda\le |\vecbox{q}|\le\Lambda$,
where $\vecbox{q}$ is the momentum of one nucleon.
One can obtain $V_L$ corresponding to $\Lambda=\Lambda_L$ by solving the WRG
equation with the initial condition $V=V_H$ for $\Lambda=\Lambda_H$.
Actually, the solution of the WRG equation, \Eq{rge}, is identical to
Feshbach's effective interaction\cite{feshbach} and is on-shell
energy dependent.

The WRG equation can also be obtained in a different way, as
has been done by Birse {\it et al.}\cite{birse}
They started with the Lippmann-Schwinger equation for $NN$-scattering
in a model space with a cutoff $\Lambda$:
\begin{eqnarray}
\eqn{lippmann}
 T^{(\alpha)}(k',k;p) &=& V^{(\alpha)}(k',k;p,\Lambda)
 + {M\over 2\pi^2}\!\! \int_0^\Lambda \!\! d\bar{k} \bar{k}^2
{V^{(\alpha)}(k',\bar{k};p,\Lambda)\: T^{(\alpha)}(\bar{k},k;p) \over p^2
- \bar{k}^2 + i\epsilon}\ .
\end{eqnarray}
Then, they differentiated both sides of the equation with respect to
$\Lambda$ and imposed the constraint that the full off-shell T-matrix is
invariant with respect to changes in $\Lambda$: {\it i.e.},
$\partial T /\partial\Lambda=0$.
This procedure leads to \Eq{rge}.
In Birse's procedure, one does not have to be careful with regard to
the choice of the reference frame in which
one is working; $p$, $k$ and $k'$ are the relative momenta of the
two-nucleon system, and $\Lambda$ is a cutoff for the relative momentum.
In case of the CM frame, the meanings of the variables in \Eq{rge}
are the same for the two derivations.
In deriving \Eq{rge} using Birse's procedure, it is essential to use
the condition that the full off-shell T-matrix is invariant.
In fact,
this condition is a natural choice if one wants to obtain an RG
equation which is the same as that obtained from the path integral.
Let us explain this point.
Integrating out the momentum shell in the path integral $Z$ does not
change the Green functions obtained from $Z$.
In non-relativistic quantum mechanics, on the other hand, the
Green function for the two-nucleon system is given by
\begin{eqnarray}
 \eqn{green}
 {1 \over E - H + i\epsilon} 
&=&{1 \over E - H_o + i\epsilon}
+{1 \over E - H_o + i\epsilon} V {1 \over E - H_o + i\epsilon}
+\cdots\\\nonumber
&=&{1 \over E - H_o + i\epsilon}
+{1 \over E - H_o + i\epsilon} T {1 \over E - H_o + i\epsilon}\ ,
\end{eqnarray}
where $H$ ($H_o$) is the full (kinetic term of the) Hamiltonian and $T$ is
the T-matrix.
If one imposes, for consistency with the path integral case, the
condition that
the Green function is invariant with respect to changes in the cutoff,
then this means that the full off-shell T-matrix is invariant, as is obvious
from \Eq{green}. This is why
full off-shell T-matrix invariance is a natural choice
to realize consistency with the path integral method.

To this point, we have
discussed the model-space reduction scheme using the WRG equation.
In deriving the WRG equation, we
started from the path integral so that we would perform the model-space
reduction in a manner consistent with the construction of the effective
Lagrangian used in NEFT.
In addition to this method, two other methods have been proposed for the
model-space reductions,
the \vlowk\ method\cite{vlowk} and the unitary transformation
(UT)\cite{UT} method.
Although these methods are not consistent with the path integral,
several authors have argued that these methods can be applied to NEFT
for model-space reduction.
For this reason, we regard these methods as possibilities for
the model-space reduction scheme in NEFT and examine their
applicability to NEFT; indeed,
it is interesting to consider the results obtained using these methods.
We briefly describe these methods in the following.

We start with the \vlowk\ method developed by Bogner {\it et al.}
\cite{vlowk}
One representation of the \vlowk\ method is the RG equation
which enforces the condition that
the half on-shell
T-matrix be independent of the cutoff.
The RG equation in this case is given by\cite{rg-vlowk}
\begin{eqnarray}
 \eqn{rg-vlowk}
 {\partial V^{(\alpha)}(k',k;\Lambda) \over\partial\Lambda} 
= {M\over 2\pi^2} V^{(\alpha)}(k',\Lambda;\Lambda){\Lambda^2\over\Lambda^2-k^2}
T^{(\alpha)}(\Lambda,k;\Lambda)\ ,
\end{eqnarray}
where the argument of the on-shell energy in $V^{(\alpha)}$ is suppressed,
because $V^{(\alpha)}$ does not depend on it if the starting potential is
independent of the on-shell energy.
The T-matrix is denoted by
$T(k',k;p)$, and the meanings of the arguments here are the same as before.
A low-momentum potential with a given cutoff can be obtained by solving
this equation. A practical method to solve this equation is discussed in 
Ref.~\citen{vlowk}.
This method gives a low-momentum potential which is non-hermitian.

Next, we consider the UT method.
We follow the procedure discussed in Ref.~\citen{UT}.
For a given value of the momentum cutoff, 
we set up a model space and its complementary space.
Then, we introduce the projection operators onto those spaces as
\begin{eqnarray}
 \eqn{eta}
\eta &=& \int {d^3 q\over (2\pi)^3} \left| \vecbox{q}\, \right\rangle \left\langle \vecbox{q}\,
  \right| \,\, ,  \quad \quad \quad
\left| \vecbox{q} \, \right| \leq \Lambda \,\, , \\
\eqn{lambda}
\lambda &=& \int {d^3 q\over (2\pi)^3} \left| \vecbox{q} \,\right\rangle  \left\langle
\vecbox{q} \, \right| \,\,  ,
\quad \quad \quad
\left| \vecbox{q} \, \right| > \Lambda \,\, .
\end{eqnarray}
Now, we derive an effective Hamiltonian acting on only the model space
by performing the unitary transformation
\begin{eqnarray}
 \eqn{transf}
{\cal H} = U^\dagger H U\ ,
\end{eqnarray}
with a condition
\begin{eqnarray}
 \eqn{bed}
\eta {\cal H} \lambda = \lambda {\cal H} \eta = 0~.
\end{eqnarray}
The low-momentum potential obtained using the UT method is defined by
\begin{eqnarray}
 \eqn{v-ut}
V \equiv \eta {\cal H} \eta - \eta H_o \eta\ .
\end{eqnarray}
The low-momentum potential obtained in this way preserves the on-shell
T-matrix elements, which has been shown in Ref.~\citen{UT}.
(For more details about the derivation and features of the UT-based
low-momentum potential, we refer the reader to Ref.~\citen{UT}.)
It is noted that the UT-based low-momentum potential
is also obtained from the non-hermitian low-momentum potential obtained with the
\vlowk\ method using the hermitization method proposed by 
Suzuki.\cite{suzuki}

\subsection{{\it Naturalness}, {\it systematicness} and {\it integrability} }
\label{sec_natural}

In NEFT, we rely on assumptions of
{\it naturalness}, {\it systematicness}, and {\it integrability}. 
These assumptions are relevant to the size of LECs and
are necessary for a convergent perturbative calculation
following a counting rule.
We explain these assumptions by considering,
for simplicity, a pionless effective field theory (\piless), where only
the nucleon is dynamical.\footnote{
In NEFT with the pion integrated out, 
chiral symmetry is not assumed because
the pion mass is considered to be very large and chiral symmetry is
not a good symmetry in that theory.
}
In \piless, all interactions are given by $NN$ contact interactions
with $2n\ (n=0,1,2\ \cdots)$ derivatives.
In an $S$-wave scattering, for example, 
\begin{eqnarray}
 V(k',k) = C_0 + C_2 (k^2+k'^2) + \cdots \ ,
\end{eqnarray}
where $C_{2n}$ is the coupling of the contact interaction with $2n$
derivatives.
All d.o.f. other than those of the nucleon have been integrated out and 
their effects are {\it assumed} to be accurately simulated by the contact
interactions. We refer to this assumption as {\it integrability}.
At leading order, one constructs a nuclear force in terms of the contact
interaction with no derivatives.
At the second leading order, the contact interactions with zero and two
derivatives are included in the nuclear force.
At the $m$-th leading order, the contact interactions with
$2n\ (n=0,1,2\ \cdots\ , m-1)$ derivatives are taken.
Because LECs of these contact interactions cannot be
determined by the assumed symmetries alone, the convergence of the
perturbation is not guaranteed from the outset.
For a convergent perturbation,
NEFT assumes {\it naturalness}, which we express by the condition
\begin{eqnarray}
 \eqn{natural}
 {C_{2(n+1)}\Lambda^2 \over C_{2n}} \ll 1\ .
\end{eqnarray}

The last assumption, {\it systematicness}, is defined as follows.
Suppose we have a set of values of LECs for a $NN$-potential
at a given order.
Then, {\it systematicness} means that when 
higher-order terms including new LECs are added to the $NN$-potential,
the values of the LECs for the lower-order terms do not change
drastically.
In other words,
an $NN$-potential {\it to be} parameterized by NEFT is assumed
to exhibit behavior that can be
parameterized by a
one- (and multi-) pion-exchange potential plus
a convergent power series of the nucleon three-momentum squared,
when we consider NEFT in which the nucleon and pion are dynamical.
In fact, as we will see, {\it systematicness} is useful criterion to
find the proper model-space reduction.

The assumptions explained above, {\it integrability}, 
{\it naturalness}, and {\it systematicness},
are not independent ideas; 
heavier d.o.f. are assumed to be
{\it integrable} to form an interaction that satisfies
{\it naturalness} and {\it systematicness}.
Also, {\it naturalness} and {\it systematicness} seem to possess the same
meaning.
If we consider very high order perturbations, this is probably true.
However, in practical cases in which we consider a few orders of perturbation,
they are not always simultaneously satisfied; we find in our results
a case in which only {\it naturalness} is satisfied, while {\it
systematicness} is not.
To this time, many authors have not clearly distinguished
these assumptions because they are not independent.

Although these assumptions seem to be reasonable,
quantitative tests of their likelihood are certainly called for.
Even though quantitative studies of NEFT have been
done extensively during the last decade, 
we wonder whether these studies are
sufficiently useful to assess the likelihood of these assumptions.
What typically has been done in previous studies is as follows.
In the case of NEFT, \veft\ is constructed and 
the LECs are determined at a given
order with the use of low-energy experimental data.
In this way, it was
found that the size of the LECs is {\it natural}.
The convergence of the perturbation was also studied
by going to higher-order perturbations.
It was found that
additional terms give
smaller contributions at higher orders.
The higher-order corrections yield better
predictions and widen the applicable energy region.
Although all these results support the assumptions,
they only amount to examinations of
a necessary condition.
The problem is that there is no way to know whether \veft\ at a
given order accurately simulates the $NN$-potential {\it to be} parameterized.
Even if the {\it naturalness} is satisfied at a given order,
the situation may change when one goes to a much higher order,
as is seen in our result given below. 
In this case, the {\it systematicness} is not realized.
This suggests that the above-mentioned
examination of (the necessary condition of) the assumptions is insufficient.
The problem arises from the fact that this {\it bottom-up} approach never
allows one to know the $NN$-potential {\it to be} parameterized by \veft.
By contrast, in our {\it top-down} approach, we know the
$NN$-potential to be parameterized,  which is obtained from \vph\ by
integrating out the d.o.f. other than those considered explicitly.
Even though we do not start from an underlying theory like QCD,
we regard the obtained potential to be parameterized by \veft\
for the following reasons:
it reproduces the low-energy $NN$-data; 
it is free from model dependence, {\it i.e.}, dependence on the nature of
the description of small scale
physics, as we see below; its behavior correctly models the
large scale physics, and the effects of the d.o.f. integrated out.
Clearly, our {\it top-down} approach provides a much better examination
of the basic assumptions of NEFT.
We know how well \veft\ at a given order simulates the $NN$-potential
{\it to be} parameterized, and therefore we can study the likelihood of
the {\it systematicness} as well as that of the {\it naturalness}.

\subsection{$NN$-potential based on nuclear effective field theory}
\label{sec_veft}

As explained in the Introduction,
we start with \vph\ and reduce its model space
to obtain the corresponding model-space potential, to which we refer
as $V_M$.
The model space is reduced up to an appropriate size for \eftpi\ (\piless).
We use the notation \eftpi\ in reference to NEFT in which the nucleon
and pion explicitly appear,
while \piless\ includes only the nucleon explicitly.
We use \vpi (\vpiless) to represent the $NN$-potential corresponding to
\eftpi (\piless). The obtained potential,
$V_M$, is simulated by \vpi\ (\vpiless)
with suitably adjusted LECs.
In this subsection,
we present expressions for \vpi\ and \vpiless\ that are used in
this work.

For \vpi, we use
a combination of the OPEP and the contact interactions
defined in a model space, $0\le|\vecbox{k}|\ (|\vecbox{k}^\prime|) \le\Lambda$,
as
\begin{eqnarray}
 \eqn{vpi}
  \bra{\vecbox{k}^\prime}V\ket{\vecbox{k}} &=&
- \vecbox{\tau}_1\cdot\vecbox{\tau}_2 \,\frac{g_A^2}{4 f_\pi^2}\,
\frac{ \vecbox{\sigma}_1\cdot \vecbox{q}\,\vecbox{\sigma}_2\cdot \vecbox{q}}
{\vecbox{q}^2+m_\pi^2}\\\nonumber
&+& C_0 + (C_2 \delta ^{ij} + D_2 \sigma^{ij})
 q^i q^j
+ (C_4 \delta^{ij}\delta^{kl} + D_4 \sigma^{ij}\delta^{kl})
 q^i q^j q^k q^l\ ,
\end{eqnarray}
with
\begin{eqnarray}
 \sigma^{ij} = \frac{3}{\sqrt{8}} \left(
\frac{\sigma_1^i \sigma_2^j + \sigma_1^j \sigma_2^i}{2}
- \frac{\delta^{ij}}{3} \vecbox{\sigma}_1 \cdot \vecbox{\sigma}_2 \right)\ .
\end{eqnarray}
The first term is the familiar OPEP, while the rest represent contact
interactions.
The momentum transferred is denoted by
$\vecbox{q}\equiv\vecbox{k}^\prime-\vecbox{k}$.
The quantities
$g_A$ and $f_\pi$ are the axial coupling constant and the pion decay
constant, respectively.
We do not include the Coulomb interaction, because we consider only
proton-neutron scattering in this work.
The LECs of the contact interactions are channel dependent.
In this work, we are concerned with the $^1S_0$ and
$^3S_1$-$^3D_1$ channels.
The $D_2$- and $D_4$-terms are relevant to the $^3S_1$-$^3D_1$ channel only.
The LECs are $\Lambda$ dependent and, in the case we use the WRG equation
[\Eq{rge}], on-shell energy dependent.
The expression for \vpiless\ is obtained by simply omitting the
OPEP from \Eq{vpi}.

The $NN$-potential [\Eq{vpi}] we use is not fully
consistent with Weinberg's counting. 
Because we consider contact interactions with zero, two, and
four derivatives,
we should include more irreducible graphs,
such as a TPEP (two-pion exchange potential) and an OPEP
with more than a single derivative $\pi NN$ coupling.
Regarding the contact interactions, 
the zero- and the two-derivative terms are the most general, as long as we are
concerned with the $^1S_0$ and $^3S_1$-$^3D_1$ channels.
However, there are other types of four-derivative terms that have not
been considered here.
Nevertheless, we use \Eq{vpi} for \vpi\ to simplify our analysis.
In spite of this simplification,
our result should not be changed essentially,
because we employ a rather small cutoff value, and therefore the
details of the TPEP play only a minor role;
the effect of the incomplete structure of the four-derivative contact
interaction is expected to be small.
Below we find
that this is indeed the case.

In order to put the $NN$-potential into the Lippmann-Schwinger
equation [\Eq{lippmann}], it is useful to have expressions
of the $NN$-potential for each channel.
Such expressions for the $NN$-potential [\Eq{vpi}] are presented in
Appendix \ref{appendix_NN}.
Numerical values of the LECs involved are also given in Appendix
\ref{appendix_NN} and in Table \ref{tab_lecs}.

\section{Result}\label{sec_result}

This section consists of three subsections.
In \S \ref{subsec_test}, we test the possible methods for the model-space
reduction scheme following the procedure described in the Introduction.
After finding an appropriate model-space reduction,
we examine the evolution of $NN$-potential
due to the reduction of the model space
in \S \ref{subsec_evolve}.
We find that the model-space potential evolves and comes to have
behavior that can be accurately simulated by contact interactions alone;
hence, the
pion-exchange potential is no longer needed.
In \S \ref{subsec_deu}, we present results for the deuteron properties and
discuss them.

\subsection{Test of model-space reduction}\label{subsec_test}
\subsubsection{Wilsonian renormalization group method}

\begin{figure}[b]
\begin{center}
 \includegraphics[width=80mm]{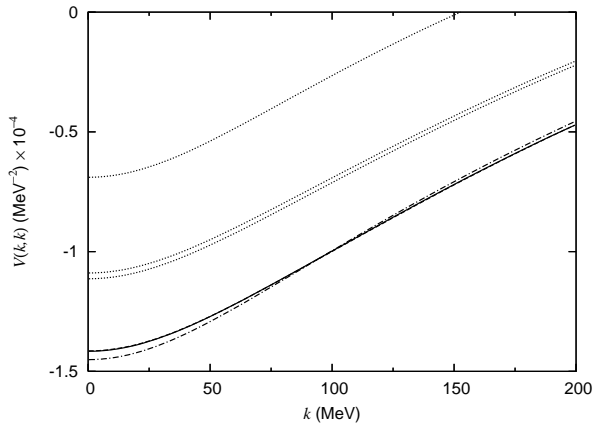}
\end{center}
\caption{\label{fig_vmm} 
The evolution of various phenomenological $NN$-potentials following the WRG
 equation. 
The diagonal momentum-space matrix elements of the potentials relevant
 to the $np$ $^1S_0$ partial wave scattering are shown.
The upper three dotted curves represent the bare phenomenological
 potentials: the upper, the Reid93, the Nijmegen I and the CD-Bonn
 potentials.
The lower three curves are the model-space potentials ($V_M$) with
$\Lambda$ = 200 MeV and $p$ = 10 MeV derived from the upper three bare potentials.
The solid, dashed and dash-dotted curves result from
the CD-Bonn, the Nijmegen I and the Reid93 potentials, respectively.
The solid and the dashed curves are nearly coincident.
}
\end{figure}

As explained in the preceding section, we start with a realistic
phenomenological $NN$-potential (\vph) and then reduce
its model space to obtain the corresponding model-space potential ($V_M$)
to be simulated by \vpi\ containing suitably adjusted LECs.
We use three \vph's:
the CD-Bonn,\cite{bonn} the Nijmegen I\cite{nij} and
the Reid93\cite{nij} potentials.
For the proton-neutron $^1S_0$ partial wave scattering,
we reduce their model spaces following the WRG equation [\Eq{rge}]
up to $\Lambda$ = 200 MeV.
The result is plotted in Fig.~\ref{fig_vmm}.
We see that
the three potentials, which are originally model dependent, are all
transformed into essentially the same model-space potential.
The slight difference between the Reid93 and the others may be
attributable to a small difference in the low-energy phase shift between
them.
For this reason, we discuss $V_M$ obtained from
the CD-Bonn potential as a representative in the following.
Although use of the WRG equation introduces an on-shell momentum
dependence, the effect is negligible for $V_M$
with $\Lambda$ = 200 MeV and $p\ltap$50 MeV.
\begin{figure}[t]
\begin{center}
 \includegraphics[width=80mm]{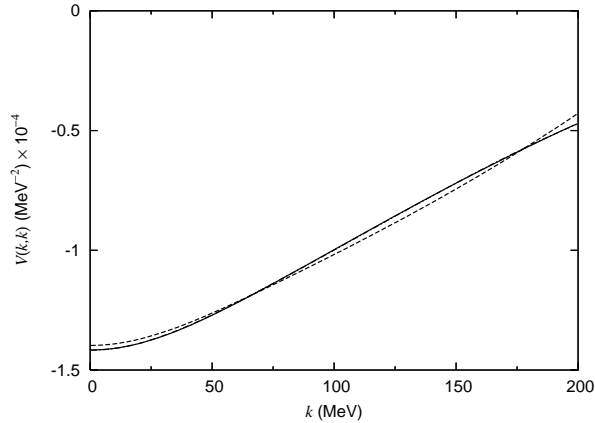}
\end{center}
\caption{\label{fig_vmm_fit} 
Simulation of the WRG-based $V_M$
for the $np$ $^1S_0$ partial wave in terms of \vpi.
The solid curve represents $V_M$ with $\Lambda$ = 200 MeV and
$p$ = 10 MeV, and it is derived from the CD-Bonn potential following the WRG
equation. The dashed curve represents \vpiI, which consists of the
OPEP and the contact interactions with zero and two derivatives.
The dash-dotted curve represents \vpiII, which consists of \vpiI\ along
 with the four-derivative contact interaction.
The couplings of the contact interactions are fixed by fitting them
to the solid curve using the least square method.
The solid and the dash-dotted curves are nearly coincident.
}
\end{figure}
In Fig.~\ref{fig_vmm_fit}, we parametrize $V_M$ in
terms of \vpi, namely, the OPEP plus the contact interactions.
Here, we use \vpiI\ to represent \vpi\ including
contact interactions with zero and two derivatives, while
\vpiII\ consists of \vpiI\ along with the
four-derivative contact interaction.
We find that \vpiI\ simulates $V_M$ rather accurately.
Using \vpiII\ yields an almost perfectly accurate simulation.
The results are consistent with the {\it systematicness} assumption of NEFT.
The numerical values of the LECs are given in Table \ref{tab_lecs}.
The numbers on the l.h.s. of \Eq{natural} are also shown. From these,
the {\it naturalness} is also found to be realized.
The phase shifts obtained with $V_M$ and \vpiII\ for this channel are in
good agreement, as seen in Table \ref{tab_phase}.
This result supports the validity of our scenario
that \veft\ and \vph\ are connected through the WRG.
\begin{table}[ht]
\renewcommand{\arraystretch}{1.6}
\caption{\label{tab_lecs}
Numerical values for the LECs involved in \veft.
The values are fitted to the model-space potential $V_M$
using the least square method.
$V_M$ is obtained from the CD-Bonn potential
 following the WRG equation.
Here, we use, for example,
7.56($-$2) to stand for
$7.56\times 10^{-2}$.
For the definition of the LECs in the first column, see
Appendix \ref{appendix_NN}.
The notation used for the \veft's in the third row is explained in the text.
}
\begin{tabular}[t]{llllllllllll}\hline\hline
\multicolumn{1}{c}{\hspace*{20mm}}& \multicolumn{4}{c|}{$\Lambda$ = 200
 MeV}&\multicolumn{4}{c}{$\Lambda$ = 70 MeV}\\
\multicolumn{1}{c}{}& 
\multicolumn{2}{c}{$p$ = 10 MeV}& \multicolumn{2}{|c|}{$p$ = 30 MeV}&
\multicolumn{2}{c|}{$p$ = 10 MeV}& \multicolumn{2}{c}{$p$ = 30 MeV}\\
&
\multicolumn{1}{c}{\vpiI}&\multicolumn{1}{c|}{\vpiII}&
\multicolumn{1}{c}{\vpiI}&
\multicolumn{1}{c|}{\vpiII}&
\multicolumn{1}{c }{\vpilessI}&
\multicolumn{1}{c|}{\vpilessII}&
\multicolumn{1}{c}{\vpilessI}&
\multicolumn{1}{c}{\vpilessII}&
\\\hline
    $C_0^{\rm (\1s0)}$(fm$^2$)&$-$5.44    &$-$5.52    &$-$5.45    &$-$5.53    &$-$10.4    &$-$10.4    &$-$11.2    &$-$11.2    \\
    $C_2^{\rm (\1s0)}$(fm$^4$)& 1.15    & 1.55    & 1.15    & 1.56    & 6.25    & 7.40    & 6.66    & 7.83    \\
$|C_2\Lambda^2/C_0|$          &0.217    &0.289    &0.217    &0.289    &7.56 ($-$2)&8.93 ($-$2)&7.46 ($-$2)&8.75 ($-$2)\\
    $C_4^{\rm (\1s0)}$(fm$^6$)&$\quad -$&$-$0.163   &$\quad -$&$-$0.164   &$\quad -$&$-$3.77    &$\quad -$&$-$3.84    \\
$|C_4\Lambda^2/C_2|$          &$\quad -$&0.108    &$\quad -$&0.108    &$\quad -$&6.41 ($-$2)&$\quad -$&6.18 ($-$2)\\
\hline                     
    $C_0^{\rm (\ts1)}$(fm$^2$)&$-$8.78    &$-$8.94    &$-$8.84    &$-$9.00    &$-$89.5    &$-$89.6    &$-$334.    &$-$334.    \\
    $C_2^{\rm (\ts1)}$(fm$^4$)& 2.11    & 2.93    & 2.13    & 2.96    & 45.0    & 48.8    & 164.    & 176.    \\
$|C_2\Lambda^2/C_0|$          &0.247    &0.337    &0.247    &0.337    &6.32 ($-$2)&6.85 ($-$2)&6.19 ($-$2)&6.61 ($-$2)\\
    $C_4^{\rm (\ts1)}$(fm$^6$)&$\quad -$&$-$0.332   &$\quad -$&$-$0.335   &$\quad -$&$-$12.5    &$\quad -$&$-$37.1    \\
$|C_4\Lambda^2/C_2|$          &$\quad -$&0.116    &$\quad -$&0.116    &$\quad -$&3.23 ($-$2)&$\quad -$&2.66 ($-$2)\\
  $D_2^{(\epsilon_1)}$(fm$^4$)&$-$0.520   &$-$0.767   &$-$0.518   &$-$0.767   & 31.6    & 36.7    & 111.    & 123.    \\
  $D_4^{(\epsilon_1)}$(fm$^6$)&$\quad -$&9.38 ($-$2)&$\quad -$&9.42 ($-$2)&$\quad -$&$-$15.2    &$\quad -$&$-$34.7    \\
$|D_4\Lambda^2/D_2|$          &$\quad -$&0.126    &$\quad -$&0.126    &$\quad -$&5.23 ($-$2)&$\quad -$&3.55 ($-$2)\\
    $C_4^{\rm (\td1)}$(fm$^6$)&$\quad -$&$-$0.254   &$\quad -$&$-$0.255   &$\quad -$& 8.99    &$\quad -$& 19.0    \\
\hline
\end{tabular}
\end{table}
\begin{table}[ht]
\renewcommand{\multirowsetup}{\centering}
\renewcommand{\arraystretch}{1.6}
\caption{\label{tab_phase}
The phase shifts obtained from various potentials for $np$-scattering.
The entries in the second row are the labels for the potentials.
The `bare' potential used here is the CD-Bonn potential, and the
model-space potential $V_M$ is obtained
 from the `bare' one following the WRG equation. The \veft's are obtained by
 fitting their LECs to $V_M$.
In the second column, $\delta$ is the phase shift, and $\epsilon_1$ is
 the mixing parameter for the $\ts1$-$\td1$ channel.
}
\begin{tabular}[t]{cccccccccccccc}\hline\hline
&\multicolumn{2}{c}{}&
\multicolumn{3}{|c|}{$\Lambda$ = 200 MeV}& \multicolumn{3}{c}{$\Lambda$
 = 70 MeV}\\
&&bare&\multicolumn{1}{|c}{$V_M$}&
\multicolumn{1}{c}{\vpiI}&\multicolumn{1}{c|}{\vpiII}&
$V_M$&
\vpilessI& \multicolumn{1}{c}{\vpilessII}\\\hline
\multirow{4}{18mm}[0mm]{$p$ = 10 MeV}
&$\delta^{(\1s0)}$& 48.04& 48.04& 52.18& 47.86& 48.04& 48.47& 48.06\\
&$\delta^{(\ts1)}$&164.47&164.47&165.22&164.46&164.47&164.42&164.47\\
&$\delta^{(\td1)}$&  0.00&  0.00&  0.00&  0.00&  0.00&  0.00&  0.00\\
     &$\epsilon_1$&  0.01&  0.01&  0.01&  0.01&  0.01&  0.02&  0.01\\
\hline
\multirow{4}{18mm}[0mm]{$p$ = 30 MeV}
&$\delta^{(\1s0)}$& 64.42& 64.42& 66.81& 64.34& 64.42& 64.70& 64.42\\
&$\delta^{(\ts1)}$&137.22&137.22&138.96&137.19&137.22&136.77&137.35\\
&$\delta^{(\td1)}$& $-$0.02& $-$0.02& $-$0.02& $-$0.02& $-$0.10& $-$0.07& $-$0.10\\
     &$\epsilon_1$&  0.24&  0.24&  0.22&  0.24&  0.24&  0.48&  0.27\\
\hline
\end{tabular}
\end{table}

\begin{figure}
\begin{center}
 \includegraphics[width=80mm]{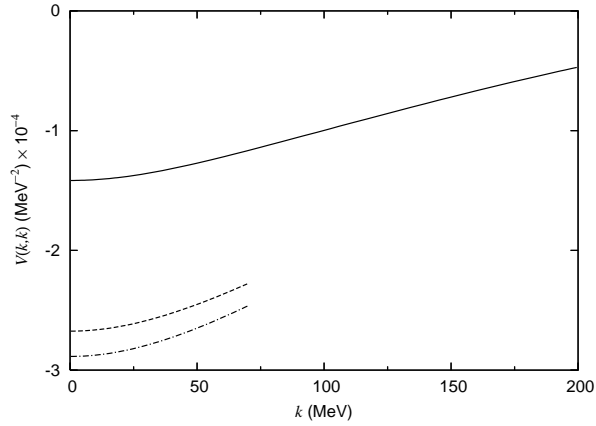}
\end{center}
\caption{\label{fig_200_70} 
The evolution of the WRG-based $V_M$
for the $np$ $^1S_0$ partial wave.
The solid curve represents $V_M$ with $\Lambda$ = 200 MeV.
The $p$ = 10 and 30 MeV cases give essentially the same solid curve.
The dashed curve represents $V_M$ with $\Lambda$ = 70 MeV and
$p$ = 10 MeV, while the dash-dotted curve corresponds to $\Lambda$ = 70
 MeV and $p$ = 30 MeV.
}
\end{figure}

\begin{figure}
\begin{center}
 \includegraphics[width=80mm]{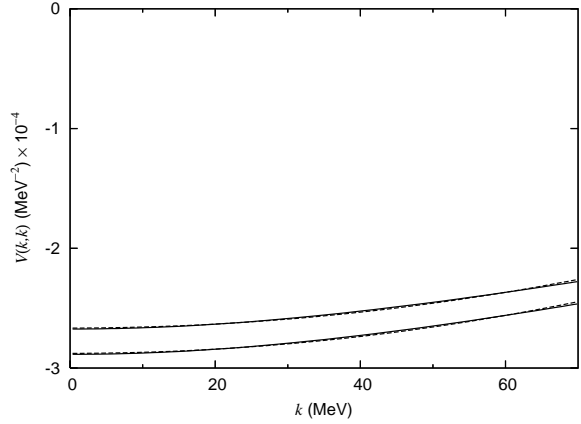}
\end{center}
\caption{\label{fig_vmm_70} 
Simulation of the WRG-based $V_M$
for the $np$ $^1S_0$ partial wave in terms of the contact interaction.
The solid curves represent $V_M$ with $\Lambda$ = 70 MeV.
The upper solid curve corresponds to $p$ = 10 MeV, while the lower one to
$p$ = 30 MeV.
Under each solid curve, there is the corresponding dashed curve.
The dashed curves represent the contact interactions with zero
and two derivatives (\vpilessI).
Their couplings are fixed following the method described in
Fig.~\ref{fig_vmm_fit}.
}
\end{figure}

\begin{figure}
\begin{center}
 \includegraphics[width=80mm]{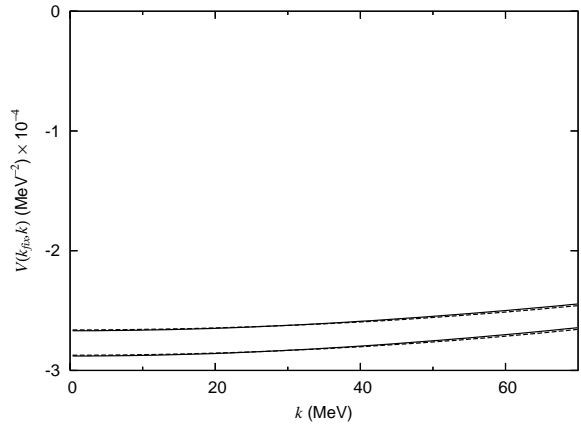}
\end{center}
\caption{\label{fig_off_diag} 
Simulation of the WRG-based $V_M$
for the $np$ $^1S_0$ partial wave in terms of \vpilessI.
Plotted here are
the {\it off}-diagonal momentum-space matrix elements of the potentials, with
$k_{fix}$ = 10 MeV. 
The other features of this figure are the same as those of
Fig.~\ref{fig_vmm_70}.
The couplings of the contact interactions in \vpilessI\ are also the
 same as those used in Fig.~\ref{fig_vmm_70}.
}
\end{figure}

We continue to reduce the model space up to $\Lambda$ = 70 MeV.
First, however, we explain why we study a case with such a small cutoff value.
One practical reason is that, as we will see, different model-space reduction
methods lead to quite different model-space potentials when we use a
small cutoff. For this reason, it is useful to study such low-momentum
potentials in order to find an appropriate model-space reduction scheme.
There is also the following formal reason.
It has been shown that a pionless EFT [that is, \piless] is useful in describing
low-energy two-nucleon systems in spite of its small model space.\footnote{
In studying a system including more than two nucleons, we have to use
a model space much larger than those for \vpiless.
Fujii {\it et al.} showed that
the binding energies of three- and four-nucleon
systems obtained from
exact calculations with a bare $NN$-potential
cannot be reproduced if one
uses the corresponding low-momentum $NN$-potential for $\Lambda\simeq$
400 MeV;
even $\Lambda\simeq$ 400 MeV is too small in this case\cite{fujii}.
The situation can be improved by using a larger model space or,
alternatively,
by considering many-body forces
generated in a model-space reduction.
}
Even though \vpiless\ results from integrating out the pion from
\vpi, no work has explicitly shown this.
In this work, we address this issue by carrying
out the following examination.
We reduce the model space for \vpi\ (or $V_M$ in the same model space)
up to a small size
and examine whether the obtained $V_M$ is well parameterized by
\vpiless.
We also examine the size of the model space appropriate for \vpiless.
(Supplementary discussion of the derivation of \vpiless\ from \vpi\ is
given in \S \ref{subsec_d0}.)
Our result is plotted in Fig.~\ref{fig_200_70}.
We adopt $p$ = 10 MeV and 30 MeV as the on-shell momentum.
We observe a large on-shell momentum dependence.
The shift of the potential due to the renormalization is larger for
$p$ = 30 MeV, which is expected from the WRG equation, \Eq{rge}.
The parametrization of $V_M$ in terms of \vpilessI\ (the contact
interactions with zero and two derivatives) is quite good, as
shown in Fig.~\ref{fig_vmm_70}.
The values of the LECs are listed in Table \ref{tab_lecs}.
We find that
the {\it off}-diagonal components of $V_M$ can also be accurately
simulated
using the contact interactions whose couplings have been fixed
by fitting to the diagonal components of $V_M$.
The result is displayed in Fig.~\ref{fig_off_diag}.
The results show that the d.o.f. integrated out,
namely the pion and the nucleon high-momentum states, can be accurately
simulated only by the contact interactions with {\it natural} couplings.
We thus find that {\it systematicness} is also realized.
This means that the WRG equation provides
an appropriate model-space reduction scheme and, simultaneously, that 
the basic NEFT assumptions discussed in \S \ref{sec_natural}
are realized.
We should recall that use of the WRG equation is also consistent with
the integration of heavier d.o.f. using the path integral, as we have
seen in \S \ref{sec_rg}.

\subsubsection{\vlowk\ method}

\begin{figure}[b]
\begin{center}
 \includegraphics[width=80mm]{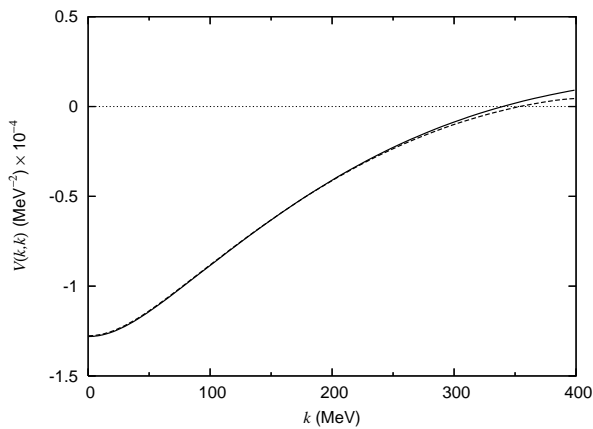}
\end{center}
\caption{\label{fig_vlowk_400} 
Comparison of the $V_M$'s obtained using the WRG and
the \vlowk\ method for the $np$ $^1S_0$ partial wave at $\Lambda$ = 400 MeV.
The solid curve corresponds to the WRG with $p$ = 30 MeV, while the
 dashed curve corresponds to the \vlowk\ method. 
In both cases, \vph\ is the CD-Bonn potential.
}
\end{figure}

\begin{figure}
\begin{center}
 \includegraphics[width=80mm]{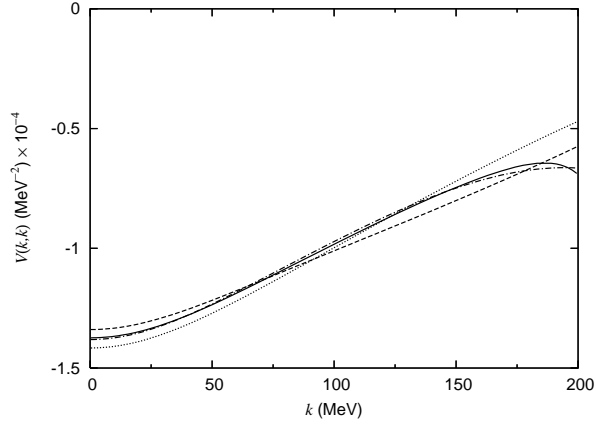}
\end{center}
\caption{\label{fig_vlowk_200} 
Simulation of $V_M$ obtained using the \vlowk\ method for the $np$
 $^1S_0$ partial wave in terms of \vpi.
The solid curve represents $V_M$
derived from the CD-Bonn potential using the
 \vlowk\ method with $\Lambda$ = 200 MeV. 
The dotted curve, shown for a comparison, is the same as the solid curve in
Fig.~\ref{fig_vmm_fit}.
The other features are the same as in Fig.~\ref{fig_vmm_fit}.
}
\end{figure}

As in the previous subsection,
we start with \vph\ and
reduce its model space, in this case using the \vlowk\ method.
Before considering our result, we briefly discuss the results obtained
with the \vlowk\ method in previous works.
In Ref.~\citen{vlowk},
it is shown in detail that
the \vlowk\ method transforms various \vph's
into essentially a single $V_M$.
In the following, therefore, we
use only the CD-Bonn potential as the starting \vph.
A simulation of the $V_M$ obtained using the \vlowk\ method 
($\Lambda\simeq$ 400 MeV)
using contact interactions is reported in Ref.~\citen{vlowk_cnt}.
In that paper, it is shown that the shifts of the potential due to the
model-space reduction ($V_M-\eta$\vph$\eta$) are accurately simulated by
the contact interactions.
We now discuss our result.
As we find in Fig.~\ref{fig_vlowk_400}, the $V_M$'s obtained from the
WRG and the \vlowk\ method are almost the same for $\Lambda$ = 400 MeV.
Even though the \vlowk\ method has no relation with the path integral
method for integrating out d.o.f., 
for this value of the cutoff,
this method is practically effective
as a model-space reduction method in NEFT.
We reduce the model space up to $\Lambda$ = 200 MeV using
the \vlowk\ method and show the resulting $V_M$
and its parameterization by \vpi\ in Fig.~\ref{fig_vlowk_200}.
We find that the
\vpi\ simulates $V_M$ rather accurately.
For momenta around the cutoff, however,
$V_M$ has a large curvature,
which cannot be simulated with the parametrization used here.

\begin{figure}
\begin{center}
 \includegraphics[width=80mm]{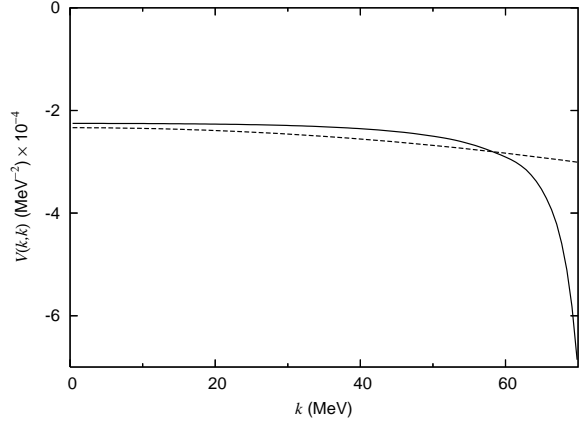}
\end{center}
\caption{\label{fig_vlowk_70} 
$V_M$ obtained using the \vlowk\ method
and an energy independent contact interaction for the $np$ $^1S_0$
 partial wave.
The solid curve represents $V_M$ with $\Lambda$ = 70 MeV 
derived with the \vlowk\ method.
The dashed curve represents the contact interactions with zero
and two derivatives.
Their couplings are fixed by fitting them
to low-energy phase shift data.
}
\end{figure}

The situation becomes worse, however, when we reduce the model space
further.
In Fig.~\ref{fig_vlowk_70}, we show $V_M$ with $\Lambda$ = 70 MeV
obtained using the \vlowk\ method.
Clearly, the simulation of $V_M$ in terms of the contact
interactions does not yield consistent behavior.
This strongly indicates that the
\vlowk\ method is not an appropriate model-space reduction in NEFT.
It should be recalled that the \vlowk\ method is not consistent with the path
integral method of integrating out the heavier d.o.f.

In case of the UT method, the result is almost the same as that
obtained with the \vlowk\ method. 
Therefore, we do not present the result for the UT method here.

\subsection{Evolution of $NN$-potential}\label{subsec_evolve}

Through the investigation carried out in the preceding subsection,
we find that the WRG equation likely provides
an appropriate model-space reduction scheme.
In this subsection, we use the WRG equation to further examine the
evolution of an $NN$-potential and thereby confirm the scenario
connecting \vph\ to \veft\ and study
the role played by the WRG equation as the method facilitating the connection.
Recall that we studied the $\1s0$ channel in the preceding subsection.
Here we study the $^3S_1$-$^3D_1$ channel.
It is interesting to study this channel, which contains a richer variety
of phenomena than
does $\1s0$; in particular,
the pion plays an important role.
This is a good place to examine the evolution of the potential and to
find an appropriate size of the model space for which the pion-exchange
potential is no longer needed.

First, we examine the evolution of various \vph's
by reducing their model spaces.
In the $^3S_1$-$^3D_1$ channel, all the \vph's used
here are phase-shift equivalent. 
Regarding the deuteron $D$-state probability ($P_D$), however, there is a
model dependence:
$P_D$(CD-Bonn) = 4.85\%,\cite{bonn} $P_D$(Nij I) = 5.66\%\cite{nij} 
and $P_D$(Reid93) = 5.70\%.\cite{nij}
It is interesting to examine whether all of these $NN$-potentials evolve
into essentially a single model-space potential, as in the $\1s0$
case, in spite of this model dependence of $P_D$.

\begin{figure}[b]
\begin{center}
 \includegraphics[width=80mm]{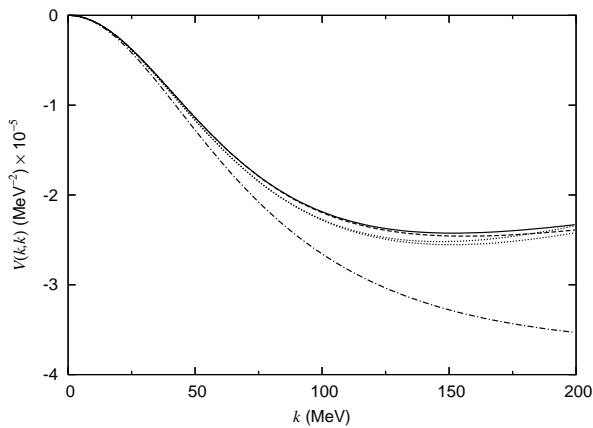}
\end{center}
\caption{\label{fig_vmm_3c1_sd} 
The evolution of $np$ $\bra{^3D_1}V\ket{^3S_1}$ for various \vph's
under the WRG equation. 
The diagonal momentum-space matrix elements are shown.
The lower two dotted curves represent the bare \vph's with 
the upper (lower) one the CD-Bonn (the Nijmegen I) potential.
The upper two curves represent $V_M$'s with
$\Lambda$ = 200 MeV and $p$ = 10 MeV derived from these two \vph's.
The solid and the dashed curves are the $V_M$'s obtained from
the CD-Bonn and the Nijmegen I potentials, respectively.
The OPEP is also shown by the dash-dotted curve.
Note that the scale of the vertical axis is different from that in
Fig.~\ref{fig_vmm} by an order of magnitude.
}
\end{figure}

\begin{figure}
\begin{center}
 \includegraphics[width=80mm]{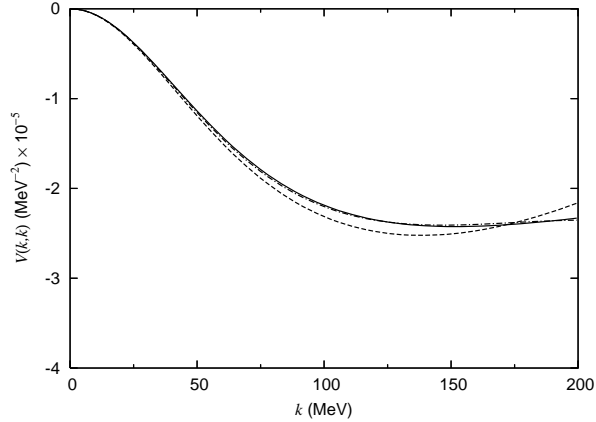}
\end{center}
\caption{\label{fig_tensor_200_cnt}
The simulation of $V_M$ for $np$ $\bra{^3D_1}V\ket{^3S_1}$ in terms of
 \vpi.
The solid curve represents $V_M$ with $\Lambda$ = 200 MeV and
$p$ = 10 MeV. It is derived from the CD-Bonn potential using the WRG
equation. 
For the $p$ = 30 MeV case, the corresponding $V_M$ is
indistinguishable in this graph from the solid curve.
The dashed curve represents \vpiI, while
the dash-dotted curve represents \vpiII.
For the notation, see the caption of Fig.~\ref{fig_vmm_fit}.
}
\end{figure}

Plots of the evolution of 
the CD-Bonn and the Nijmegen I potentials for $\bra{^3D_1}V\ket{^3S_1}$
are presented in Fig.~\ref{fig_vmm_3c1_sd},
where we use $\Lambda$ = 200 MeV and $p$ = 10 MeV.
The OPEP is also shown there.
We do not plot the evolution of the Reid93 potential, in order to make
the figure clearer.
Including the Reid93 potential would not change the discussion below.
As we see in the figure,
the initial model dependence is small for low-momentum components
because of the dominance of the OPEP tensor force.
It is noted that there is a strong model dependence in momentum components
much larger than those shown in the figure.
The shift of the potential due to the model-space reduction
is not large for our choice of $\Lambda$ and $p$, and
the degree of the model dependence remains small.
Therefore we use the $V_M$ obtained from the CD-Bonn potential as a representative 
in the following.
We also observe from the figure
that the shape of the original and the model-space potentials are
largely governed by the OPEP. 
Obviously, we need the OPEP to parameterize $V_M$ for $\Lambda$ = 200 MeV.
The simulation of $V_M$ in terms of \vpi\
is quite accurate, as seen in
Fig.~\ref{fig_tensor_200_cnt}.
The values of the LECs for this \vpi\ 
are listed in Table \ref{tab_lecs},
while the phase shifts are listed in Table \ref{tab_phase}.

\begin{figure}
\begin{center}
 \includegraphics[width=80mm]{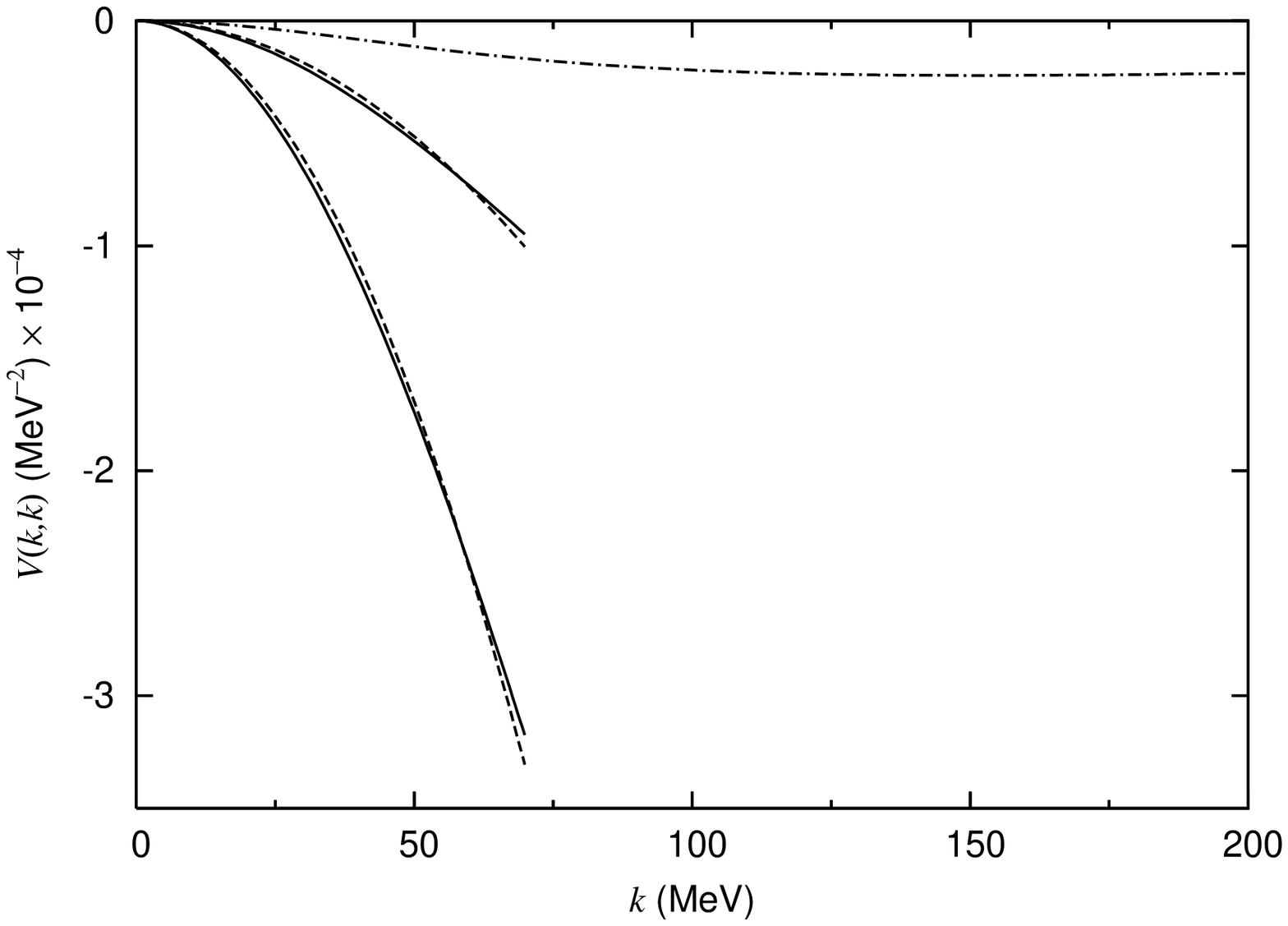}
\end{center}
\caption{\label{fig_tensor_70}
Evolution of $NN$-potential under the WRG equation
for $np$ $\bra{^3D_1}V\ket{^3S_1}$
and the simulation of $V_M$ in terms of \vpiless.
The solid curves represent $V_M$'s with $\Lambda$ = 70 MeV;
the upper solid curve corresponds to the $p$ = 10 MeV case, while the lower
one corresponds to $p$ = 30 MeV.
Each of the solid curves is simulated by the $D_2^{(\epsilon_1)}$-term alone
(\vpilessI), as shown by the accompanying dashed curve.
By including the $D_4^{(\epsilon_1)}$-term (\vpilessII),
the simulation of the solid curves
cannot be distinguished from $V_M$ in this graph.
For comparison, $V_M$ with $\Lambda$ = 200 MeV is also
plotted by the dash-dotted curve. This curve is the same as the solid
 curve in Fig.~\ref{fig_tensor_200_cnt}.
}
\end{figure}

In order to simulate a $V_M$ without the OPEP, 
what is an appropriate value of $\Lambda$?
Because the OPEP can be expanded in terms of the contact interactions as
$q^2/(q^2+m_\pi^2)\sim q^2/m_\pi^2 - q^4/m_\pi^4 + \cdots$,
the expansion is convergent for $\Lambda\ltap m_\pi/2$.
For a rapid convergence, a much smaller cutoff value is necessary.
Therefore, we reduce the model space to have the potential with
$\Lambda$ = 70 MeV $\sim$ $m_\pi/2$.
The result is shown in Fig.~\ref{fig_tensor_70} for $p$ = 10 and 30 MeV.
We find that
there is a large on-shell energy dependence
between the $p$ = 10 and 30 MeV cases,
while no difference can be discerned in the graph for $\Lambda$ = 200 MeV.
A simulation of $V_M$ in terms of the $D_2^{(\epsilon_1)}$-contact
term alone [\vpilessI; see \Eq{vpi_eps}
for the definition of the $D_2^{(\epsilon_1)}$-term]
is also shown in the figure.
Inclusion of the $D_4^{(\epsilon_1)}$-term (\vpilessII) makes the
simulation sufficiently good that no discrepancy can be seen in this graph.
The simulation is, even with this relatively large cutoff value, significantly
better than what we naively expected from the above considerations.
We can explain this as follows.
As the model space is reduced, the potential becomes so strongly
renormalized that the {\it bare} OPEP plays only a minor role; that is, the bare
OPEP contribution is hidden by
the shift of the potential due to
the renormalization.
The shift of the potential has a shape suitable for a contact
interaction expansion. 
This is the reason that the expansion is effective for a relatively large cutoff
value.
This accurate simulation supports the {\it systematicness} assumption and the values
of the LECs in
Table \ref{tab_lecs} indicate that {\it naturalness} is also
realized.

\begin{figure}[t]
\begin{center}
 \includegraphics[width=80mm]{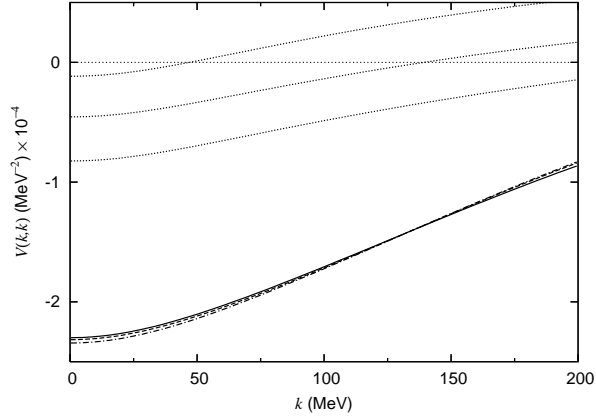}
\end{center}
\caption{\label{fig_vmm_3c1_ss} 
The evolution of $np$ $\bra{^3S_1}V\ket{^3S_1}$ for various
\vph's following the WRG equation. 
The upper three dotted curves represent the bare \vph's corresponding to
(from top to bottom) the Reid93, the Nijmegen I and the CD-Bonn
 potentials.
The lower three curves represent $V_M$'s with
$\Lambda$ = 200 MeV and $p$ = 10 MeV derived from the upper three bare potentials.
The solid, dashed and dash-dotted curves represent $V_M$'s derived from
the CD-Bonn, the Nijmegen I and the Reid93 potentials, respectively.
}
\end{figure}

It is interesting to examine the evolution of $\bra{^3S_1}V\ket{^3S_1}$
driven by the model-space reduction,
because much more significant renormalization of the potential is expected
on the basis of the following speculation.
The tensor force is known to excite a low-energy $^3S_1$ state to
a high-energy $^3D_1$ state, and vice versa.
Therefore, a sequence of transitions like
$^3S_1\rightarrow{}^3D_1\rightarrow{}^3S_1$
is renormalized into the central force of the $^3S_1\rightarrow{}^3S_1$
transition after integrating out the high-momentum states.
Below we show that this is indeed the case.
We start with three potentials
(CD-Bonn, Nijmegen I and Reid93)
for $\bra{^3S_1}V\ket{^3S_1}$.
In this case, the initial model dependence is clear for the low-momentum
matrix elements, as seen in Fig.~\ref{fig_vmm_3c1_ss}.
\begin{figure}
\begin{center}
 \includegraphics[width=80mm]{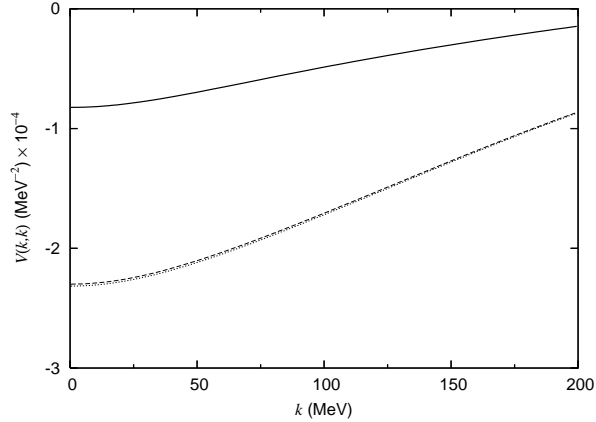}
\end{center}
\caption{\label{fig_3c1_ss_200}
Evolution of $np$ $\bra{^3S_1}V\ket{^3S_1}$
under the WRG equation. 
The solid curve represents the original CD-Bonn potential while the dashed
(dotted) curve
is the corresponding $V_M$ with $\Lambda$ = 200 MeV and
$p$ = 10 MeV ($p$ = 30 MeV).
}
\end{figure}
We see from the figure that they
evolve into essentially the same $V_M$.
From this point, we use the $V_M$ from the CD-Bonn potential as a
representative.
Our expectation of a strong renormalization is confirmed by comparing
Fig.~\ref{fig_vmm_3c1_ss} with Fig.~\ref{fig_vmm_3c1_sd}.
Because of this strong renormalization, 
the on-shell energy dependence of $V_M$ is slightly discernible even at
$\Lambda$ = 200 MeV (Fig.~\ref{fig_3c1_ss_200}).
We do not show the result but mention that this $V_M$ is
accurately simulated by \vpi\
at the same level found in Fig.~\ref{fig_vmm_fit}.
When we reduce the model space up to $\Lambda$ = 70 MeV, we observe
a significant
evolution, as seen in Fig.~\ref{fig_3c1_ss_70}.
\begin{figure}
\begin{center}
 \includegraphics[width=80mm]{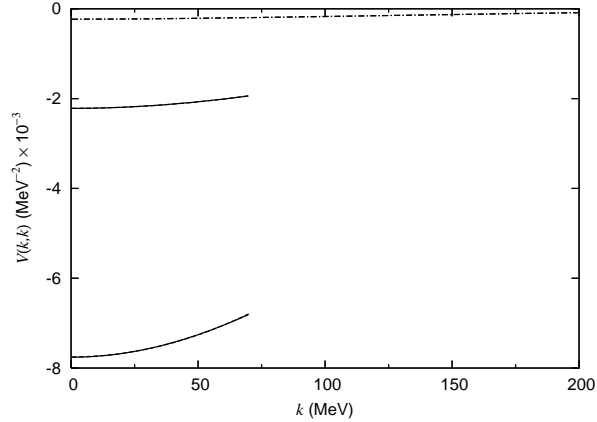}
\end{center}
\caption{\label{fig_3c1_ss_70}
Evolution of $V_M$ under the WRG equation
for $np$ $\bra{^3S_1}V\ket{^3S_1}$
and the simulation of $V_M$ with \vpiless.
The solid curves represent the model-space potentials with $\Lambda$ = 70 MeV;
the upper solid curve corresponds to the $p$ = 10 MeV case, while the lower
one corresponds to $p$ = 30 MeV.
Each of the solid curves is simulated using the contact
interactions with zero and two derivatives (\vpilessI),
as shown by the accompanying
dashed curve; each of the dashed curves is almost completely indistinguishable from
the corresponding solid curve in this graph.
For comparison, $V_M$ with $\Lambda$ = 200 MeV is also
plotted by the dash-dotted curve. This curve is the same as the dashed
 curve in Fig.~\ref{fig_3c1_ss_200}.
Note that the dashed and the dotted curves in
Fig.~\ref{fig_3c1_ss_200} are indistinguishable
with the scale used in this figure.
}
\end{figure}
Even with this evolution, 
we confirm again that the {\it naturalness} and the {\it systematicness}
are maintained,
as seen from Fig.~\ref{fig_3c1_ss_70} and Table \ref{tab_lecs}.
Although we do not present the result, we mention that
the contact interactions determined by fitting to the diagonal
components also accurately reproduce the off-diagonal momentum components.
We do not present the result for the evolution of
$\bra{^3D_1}V\ket{^3D_1}$ either, but we find that the
trend of the result is the same as that in the $\bra{^3D_1}V\ket{^3S_1}$ case.

As we have shown, the large model dependence of the original \vph's
for the $\ts1$-$\td1$ channel have
disappeared after the model-space reduction under the WRG equation,
in spite of the differences in $P_D$ among the original \vph's.
This is not surprising, however, because these differences in $P_D$ are
due solely to modeling of small scale phenomena.
This point will be discussed in the next subsection.
We have also shown that
\vpi\ (\vpiless) with suitably adjusted LECs accurately simulates
the low-momentum $V_M$ obtained using
the WRG equation and reproduces
the phase-shifts, as shown in Table \ref{tab_phase}.

\subsection{Deuteron in nuclear effective field theory}
\label{subsec_deu}

We now examine how well \vpi\ (\vpiless) reproduces the wave function
and the properties of the deuteron.
For this purpose, we derive $V_M$ from \vph\
using the WRG equation, where the
on-shell energy is set to the deuteron binding energy (B.E.).
We obtain \vpi\ (\vpiless) by fitting the LECs involved to $V_M$.
We solved the model-space deuteron eigenvalue problem
in the momentum space
with the use of $V_M$ ($\Lambda$ = 200, 70 MeV), \vpi\ ($\Lambda$ = 200 MeV)
and \vpiless\ ($\Lambda$ = 70 MeV).
The deuteron B.E. obtained in each case is listed
in Table \ref{tab_deu}.
We find that
\vpiII\ and \vpilessII\ reproduce the B.E. quite well.
\begin{table}[t]
\renewcommand{\multirowsetup}{\centering}
\renewcommand{\arraystretch}{1.4}
\tabcolsep=2.3mm
\caption{\label{tab_deu}
The deuteron binding energy (B.E.) and the $D$-state probability ($P_D$)
obtained from various potentials.
In the second column, appear the \vph's from which the $V_M$'s are obtained
using the WRG equation.
The numbers in the third column are not from the original paper but from
 our numerical calculation.
We use the extended definition of $P_D$ [\Eq{pd}] here.
}
 \begin{tabular}[t]{clccccccc}\hline\hline
&&\multicolumn{1}{c}{}& \multicolumn{3}{|c|}{$\Lambda$ = 200 MeV}&\multicolumn{3}{c}{$\Lambda$ = 70 MeV}\\
&& \multicolumn{1}{c|}{bare} & $V_M$& \vpiI& \vpiII& \multicolumn{1}{|c}{$V_M$}& \vpilessI& \vpilessII\\\hline
                                  &CD-Bonn& 2.224& 2.224& 2.509& 2.225& 2.223& 2.230& 2.223\\
\multirow{3}{15mm}[5.4mm]{B.E.\\(MeV)}&Nij I& 2.226& 2.226& 2.507& 2.227& 2.225& 2.232& 2.224\\
                                   &Reid93& 2.225& 2.224& 2.538& 2.233& 2.223& 2.231& 2.223\\
\hline
                                  &CD-Bonn& 4.85& 1.44& 1.55& 1.45& 0.09& 0.07& 0.09\\
\multirow{3}{15mm}[5.5mm]{$P_D$\\(\%)}&Nij I& 5.68& 1.47& 1.58& 1.47& 0.09& 0.07& 0.09\\
                                   &Reid93& 5.70& 1.45& 1.56& 1.44& 0.08& 0.07& 0.08\\
\hline
 \end{tabular}
\end{table}

\begin{figure}[b]
\begin{center}
 \includegraphics[width=80mm]{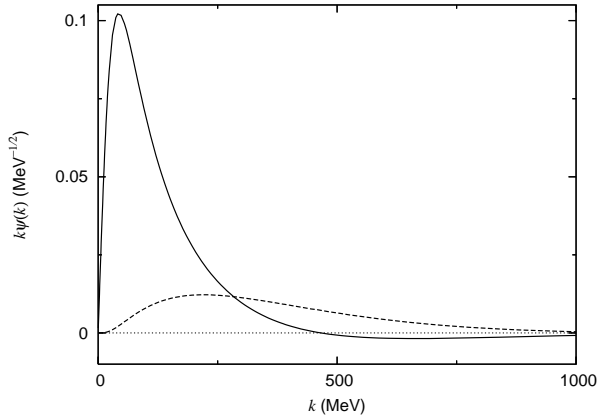}
\end{center}
\caption{\label{fig_deuteron} 
The radial part of the deuteron wave function in the momentum space.
The solid curve represents the $S$-wave, while the dashed curve represents the $D$-wave.
The vertical axis corresponds to the radial wave function multiplied by the
momentum $k$.
}
\end{figure}

How about the other deuteron properties which are often discussed in
the literature? They are, for example,
the asymptotic $S$-wave normalization $A_S$, the $D/S$-ratio $\eta$ and
the $D$-state probability $P_D$.
Actually, a calculation of these quantities is beyond the capability of
a theory like NEFT, in which a relatively small model space is used.
In order to determine
these quantities, we need information about {\it details} of the small scale
physics, which have been integrated out in NEFT. Allow us to explain
this point further.
In order to calculate these quantities, the high-momentum
components of the deuteron wave function are necessary.
The WRG equation provides a relation between the wave functions of the
full space and the model space; that is,
the WRG equation does not change the off-shell T-matrix, and
therefore the model-space wave function is obtained from the corresponding 
full-space one by just cutting off the momentum components
higher than a given cutoff.
Thus, no information is available from NEFT
about the high-momentum components of the wave function.
This is why some deuteron properties cannot be obtained in NEFT.
Of course, if we use a sufficiently large model space, the quantities can be
calculated with a good approximation.
However, as found from Fig.~\ref{fig_deuteron}, 
such a good approximation requires $\Lambda\gtap$1 GeV, which is too large for NEFT
including only the nucleon and the pion explicitly.

\begin{figure}
\begin{minipage}[t]{65mm}
 \includegraphics[width=65mm]{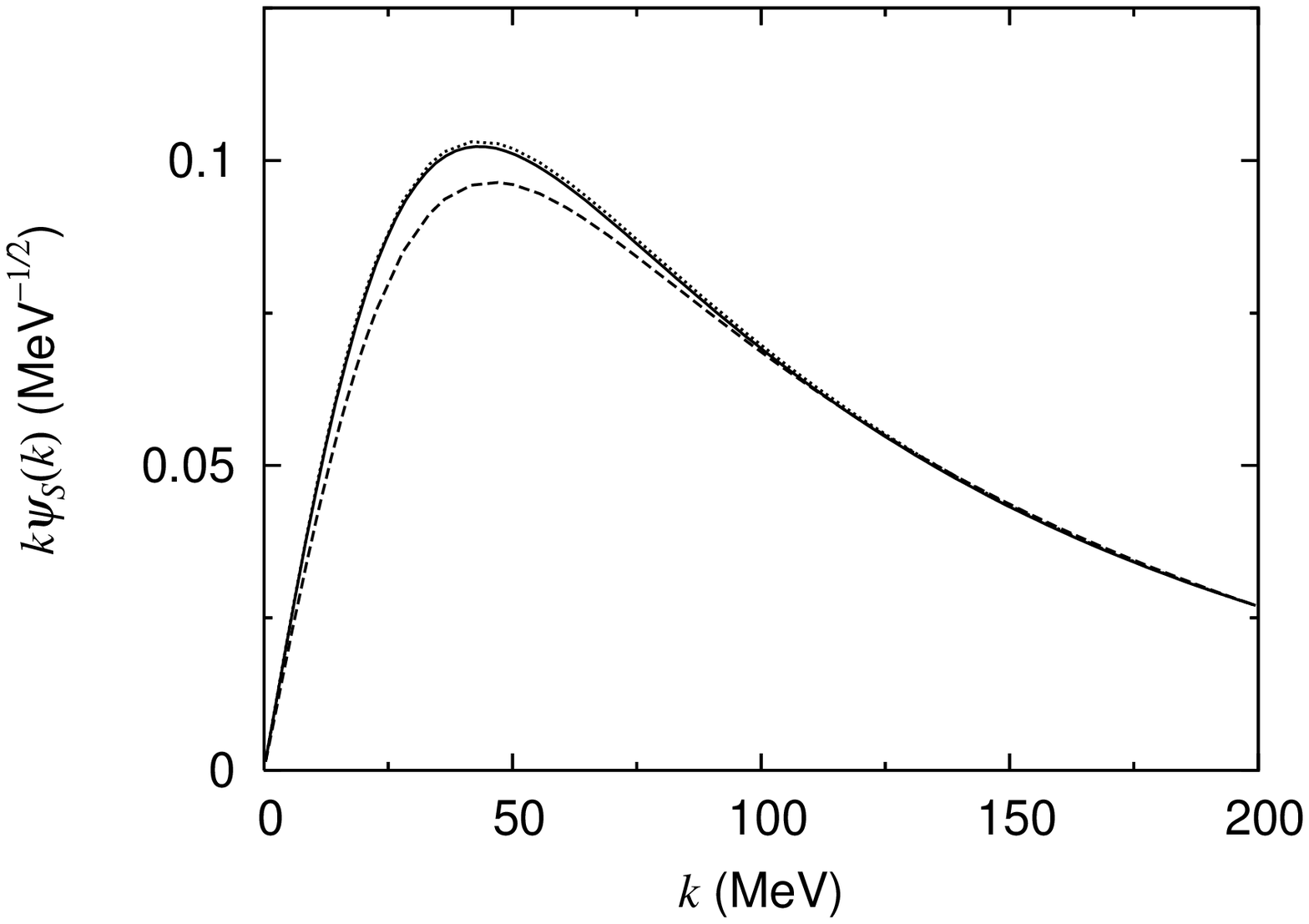}
 \caption{\label{fig_deu_L200_s}
The radial part of the deuteron $S$-wave functions in the momentum space.
The solid, dashed and dotted curves are obtained with
$V_M$, \vpiI\ and \vpiII, respectively.
Regarding the normalization of the wave functions, see the text.
 }
\end{minipage}
\hspace{5mm}
\begin{minipage}[t]{65mm}
 \includegraphics[width=65mm]{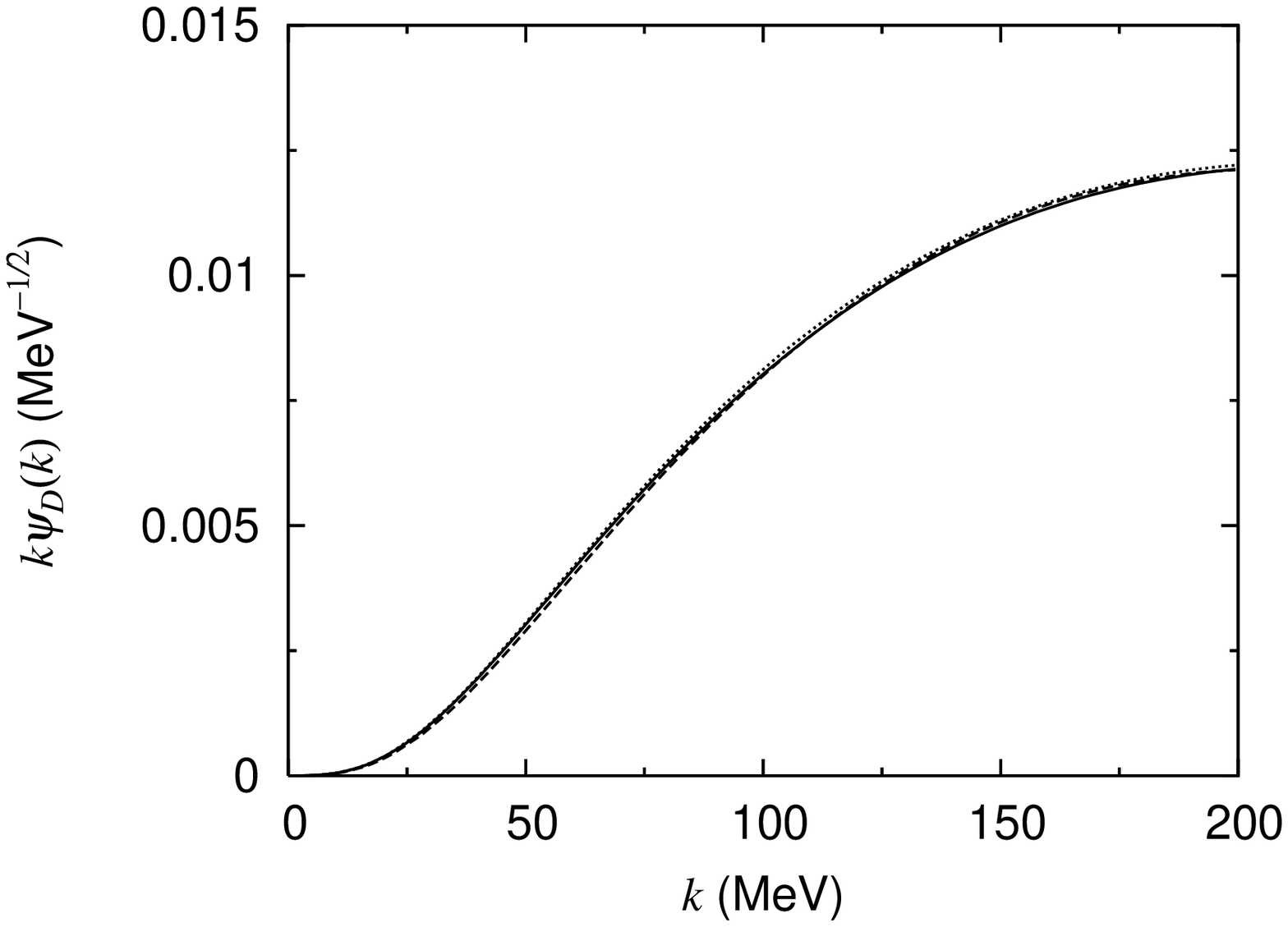}
 \caption{\label{fig_deu_L200_d}
The radial part of the deuteron $D$-wave functions in the momentum space.
The other features are the same as those in Fig.~\ref{fig_deu_L200_s}.
 }
\end{minipage}

\begin{minipage}[t]{65mm}
 \includegraphics[width=65mm]{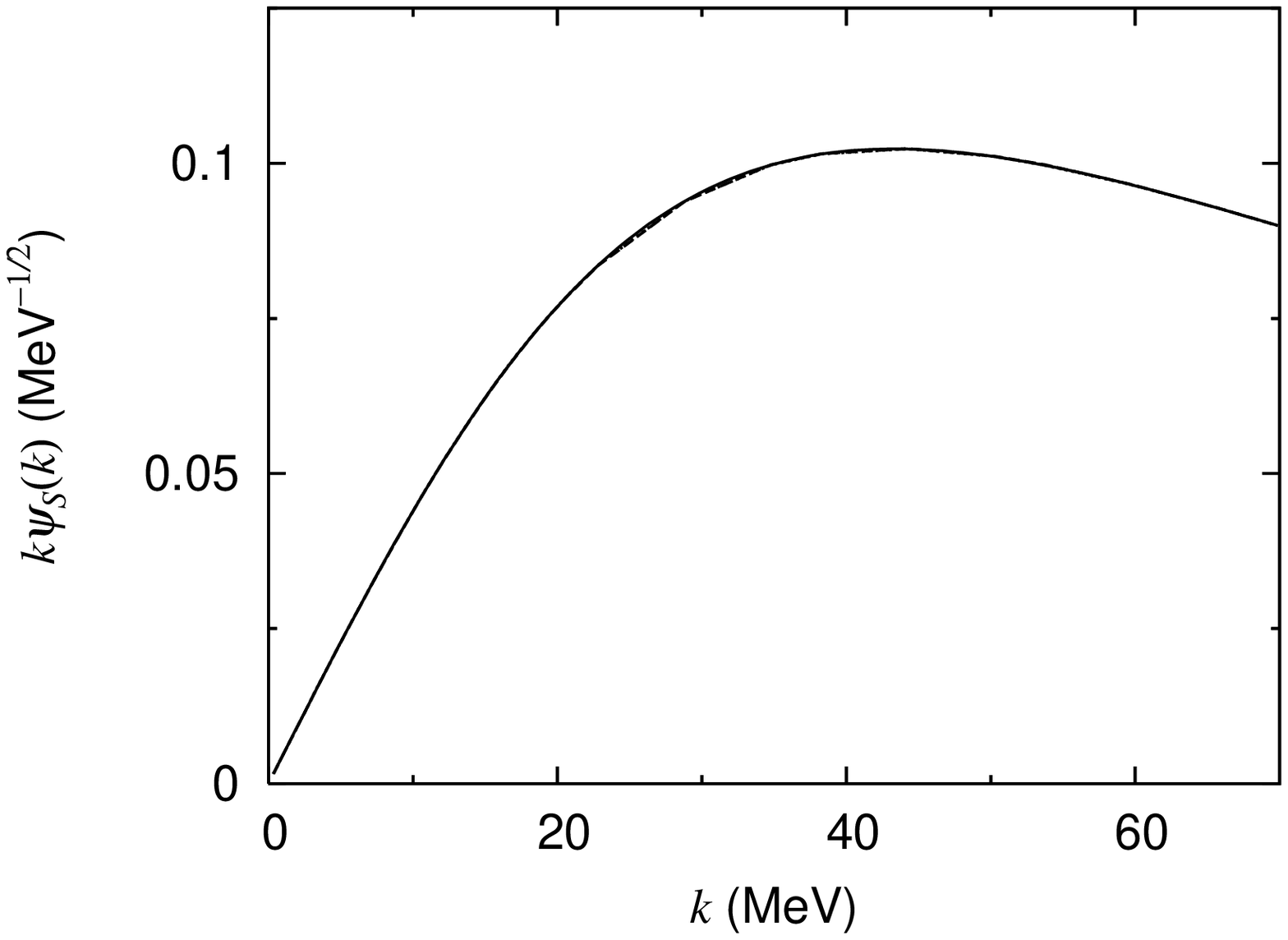}
 \caption{\label{fig_deu_L70_s}
The radial part of the deuteron $S$-wave functions in the momentum
 space. 
The solid, dashed and dotted curves are obtained with
$V_M$, \vpilessI\ and \vpilessII, respectively.
}
\end{minipage}
\hspace{5mm}
\begin{minipage}[t]{65mm}
 \includegraphics[width=65mm]{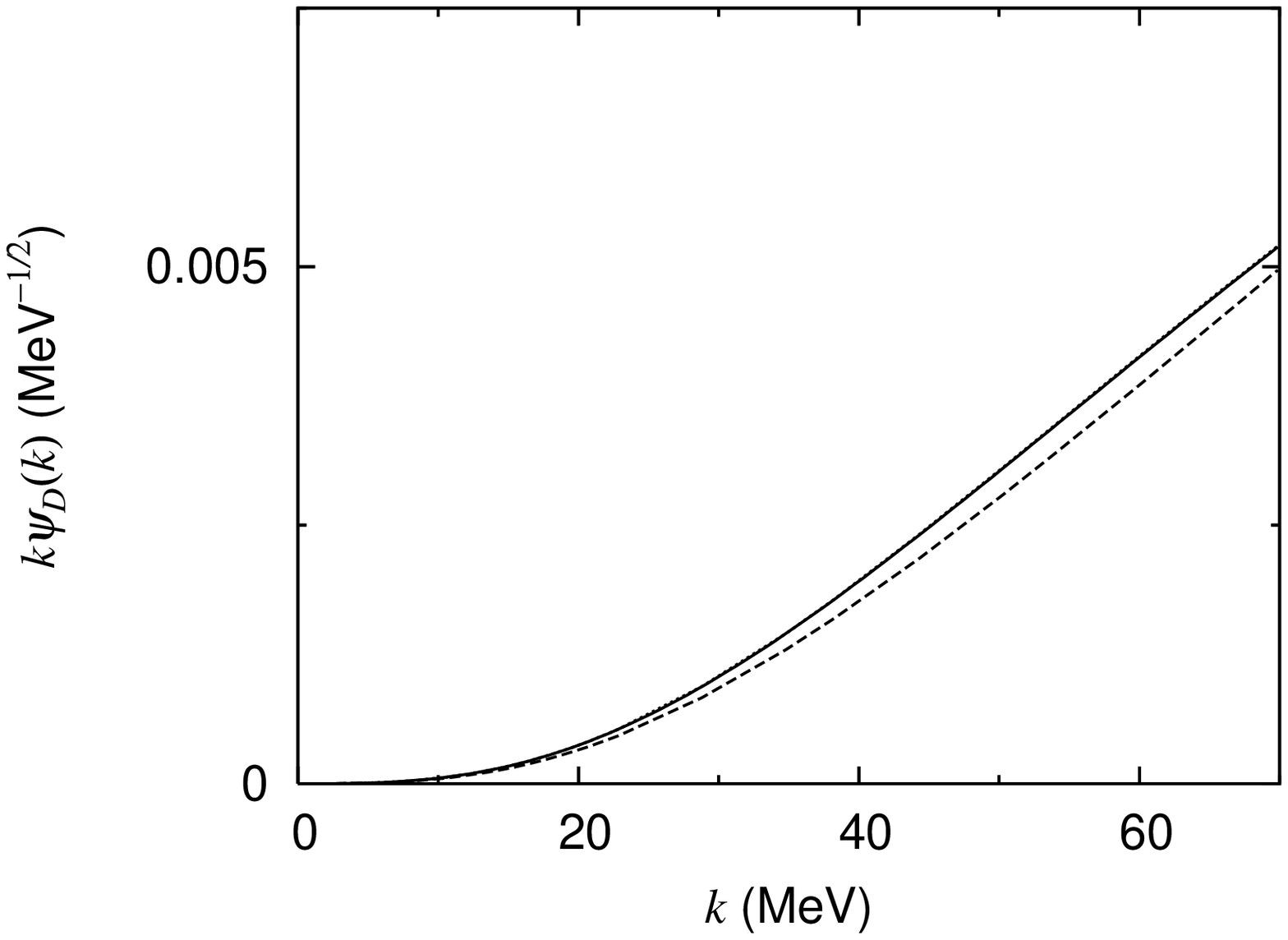}
 \caption{\label{fig_deu_L70_d}
The radial part of the deuteron $D$-wave functions in the momentum space.
The other features are the same as those in Fig.~\ref{fig_deu_L70_s}.
}
\end{minipage}
\end{figure}

Although many deuteron properties are not described in NEFT, we consider the
deuteron wave function in a model space to see how well \veft\ works;
the result is given in Figs. \ref{fig_deu_L200_s}--\ref{fig_deu_L70_d}.
In fact, the normalization of the deuteron wave function also
requires information about details of the small scale physics;
NEFT alone cannot normalize the wave function. Here, we normalized the
model-space deuteron wave function such that the normalized $S$-wave
and the wave function obtained with \vph\ give
the same amplitude at $k=\Lambda$.
We confirmed that
the wave function obtained with $V_M$ is the same as that obtained with the
corresponding \vph\ (the CD-Bonn potential in this case), as should be the case.
From the figures, we see that \vpi\ and \vpiless\ are successful in
generating the wave function.

Lastly, we discuss the deuteron $D$-state probability, $P_D$.
As we have seen, various \vph's evolve into a single $V_M$
under the WRG equation. Therefore,
if we extend 
the definition of $P_D$ to the model-space case using
\begin{eqnarray}
 \eqn{pd}
  P_D &=&{\int_0^\Lambda dk k^2 |\psi_D(k)|^2\over
\int_0^\Lambda dk k^2 \left(|\psi_S(k)|^2 +|\psi_D(k)|^2\right)}\ ,
\end{eqnarray}
the model dependence of $P_D$ disappears after the model-space
reduction.
In Table \ref{tab_deu}, we list the extended $P_D$'s for $V_M$'s
originally from various \vph's.
Actually, this result is expected by observing the full-space
deuteron wave function for the following reasons: 
(1) only weak model dependence on the choice of \vph\
is found in the low-momentum components of
the deuteron wave function;
(2) the model-space deuteron wave function is obtained from the
full-space one by
simply cutting off the momentum components higher than $\Lambda$.
We can conclude that the model dependence of $P_D$ comes from 
the details of the modeling of the small scale physics.
Coraggio {\it et al.} also discussed the point that the model dependence of $P_D$
disappears after integrating out the high-momentum states with the use
of the \vlowk\ method.\cite{coraggio}

\section{Discussion}\label{sec_discussion}

All the results presented in the previous section support our scenario,
discussed in the Introduction, for
understanding \veft\ and \vph\ in a unified manner with the help of the WRG.
In the following,
we take the viewpoint offered by this scenario to understand the relation between \veft\ and \vph.
In \S \ref{subsec_d0},
we discuss, from this viewpoint, the relation between \eftpi\ and
\piless, starting from the chiral effective Lagrangian for the pion and
the nucleon.
We also note that the WRG equation indicates
properties which \veft\ should possess.
We refer to a \veft\ possessing such properties
as a {\it proper} \veft.
However, some of these properties
have not yet been fully recognized.
In \S \ref{subsec_d1},
we discuss the expected nature of \veft\ based on consideration
of the WRG equation.
From the same viewpoint,
we also discuss the nature of a previously
constructed \veft\ in \S \ref{subsec_d2}.
We refer to such a \veft\ as a {\it conventional} \veft.
Finally, in \S \ref{subsec_d3},
we discuss what we can learn about a
description of a low-energy two-nucleon system 
based on a traditional model approach and that based on 
NEFT from the point of view developed in this work.

\subsection{From \eftpi\ to \nopiy}
\label{subsec_d0}
In \eftpi, we start with the path integral
\begin{eqnarray}
\eqn{z_lchi}
 Z &=& \int {\cal D}N{\cal D}N^\dagger {\cal D}U\
e^{i\int  d^4\!x\ {\cal L}_{\chi}}\ , 
\end{eqnarray}
where $N$ is the nucleon field, $U$ contains the pion field, and
${\cal L}_{\chi}$ is an effective chiral Lagrangian containing the
nucleon and the pion fields.
The usual procedure employed in NEFT is to specify a set of irreducible
graphs from ${\cal L}_{\chi}$ following a counting rule and to regard the
set of the graphs as the $NN$-potential or the nuclear operator.
One puts these operators into the Schr\"odinger equation and solves it to
obtain physical observables.
Because one regards a set of irreducible graphs as the nuclear
operator, the theory is no longer a quantum field theory but, rather,
non-relativistic quantum mechanics.
Here, we describe the above standard NEFT procedure in the language of quantum
field theory.

The standard procedure to calculate the $NN$-scattering amplitude
in quantum field theory is as follows:
(1) calculate the nucleon four-point Green function, $G_4$,
using the path integral of \Eq{z_lchi}; (2) apply the LSZ
reduction formula to $G_4$ to obtain the S-matrix of the
$NN$-scattering.
This procedure is not equivalent to that used in NEFT.
With the
following modification, the two procedures become equivalent.
In calculating  $G_4$ perturbatively with the use of ${\cal L}_{\chi}$,
one uses a ``counting rule'';
that is, one retains only the ladders of a set of irreducible graphs.
The set of irreducible graphs from ${\cal L}_{\chi}$
is, of course, what the counting rule specifies in the standard NEFT
procedure.
We use $\tilde{G}_4$
to represent the nucleon four-point Green function to which the
counting rule has been applied.
By applying the LSZ reduction formula to $\tilde{G}_4$,
we can obtain the S-matrix for the $NN$-scattering.
In this way, we explained the standard NEFT procedure 
in the language of quantum field theory.

Now we consider how to obtain \piless\ from \pieft.
If we restrict ourselves to a two-nucleon system and need to calculate
$\tilde{G}_4$ rather than $G_4$, it is equivalent to using the path
integral
\begin{eqnarray}
 \eqn{z_ln_2}
 Z &=& \int {\cal D}N{\cal D}N^\dagger e^{i\int  d^4\!x\ {\cal L}_N}\ ,
\end{eqnarray}
with the structure of ${\cal L}_N$ given in \Eqss{lagn}{lagn2}.
The matrix element of the interaction Lagrangian contained in ${\cal L}_N$
between free two-nucleon states is, up to an overall sign, nothing more
than the set of irreducible graphs which the counting rule specifies.
Somehow, the pion has been integrated out.
Once a path integral of the form of \Eq{z_ln_2} is obtained,
the nucleon high-momentum states can be reduced
up to a size appropriate for \piless\ following the
procedure discussed in Appendix \ref{app_rg}.
Then one obtains the effective
Lagrangian to be directly compared with the pionless Lagrangian;
the LECs involved in \piless\ can be fixed by the comparison.
In this way, one obtains \piless\ starting from an effective chiral
Lagrangian of \eftpi.

It is noted that 
we do not perform a rigorous procedure, which is far from trivial,
to obtain \Eq{z_ln_2} from \Eq{z_lchi}
by integrating out the pion.
Nevertheless, as long as we are concerned with $\tilde{G}_4$,
our procedure does make sense.

\subsection{Characteristics of $NN$-potential {\it properly} based on
  effective field theory}
\label{subsec_d1}

First, we discuss the on-shell energy dependence of \veft.
As we have seen in the result, 
\veft\ has the on-shell energy dependence.
The dependence is strong, in particular, in the case of a small value of
the cutoff.
This is a natural consequence of integrating out the nucleon
high-momentum states by use of the WRG equation.
In spite of this fact, the procedure employed in previous works to construct
\veft\ is as follows.
The irreducible graphs from a Lagrangian up to a given order
are simply multiplied by a cutoff function,
and this is used as \veft.
The high-momentum states which are not considered
explicitly in this way are assumed to be absorbed by on-shell
energy {\it independent} contact interactions.
Possible energy dependence is considered to be eliminated by using the
equation of motion.
However, as discussed by Birse {\it et al.},\cite{birse} the WRG equation
indicates
that the on-shell energy dependence introduced by the WRG equation is not
eliminated by the equation of motion.
This can also be seen clearly from the result
({\it e.g.}, see Fig.~\ref{fig_200_70})
in the previous section.
Birse {\it et al.} treated the on-shell energy dependence coming from
the nucleon high-momentum states that have been integrated out.
However, they were interested in the behavior of the potential around the fixed
point, and therefore they did not study how important the on-shell energy
dependence of \veft\ is.

We showed that \veft\ is largely on-shell energy dependent, in
particular, in the case of a small value of the cutoff.
Actually,
the large on-shell energy dependence is due to the fact that separation of
scale is not so strong.
In order for contact interactions to accurately simulate the pion contribution,
we have to take the cutoff value to be maximally $\Lambda\sim$ 70 MeV.
On the other hand, we hope \piless\ to be effective for $p\ltap$70 MeV.
Therefore, we find $p\ltap\Lambda$.
The WRG equation [\Eq{rge}] immediately indicates that there must be a
large on-shell energy dependence in such a case. By contrast,
the WRG equation tells us that
the on-shell energy dependence is negligible in the case $p\ll\Lambda$.

There is another source of on-shell energy dependence of \veft.
The on-shell energy dependence
due to the recoil correction has been
considered in a previous work.\cite{kolck-ptl}
This effect is a higher-order correction and,
as discussed in Ref.~\citen{epelbaum}, it
may be eliminated when one uses a unitary
transformation method in constructing \veft.
This type of on-shell energy dependence is not as important as that
from the model-space reduction.

\begin{figure}
\begin{minipage}[t]{65mm}
 \includegraphics[width=65mm]{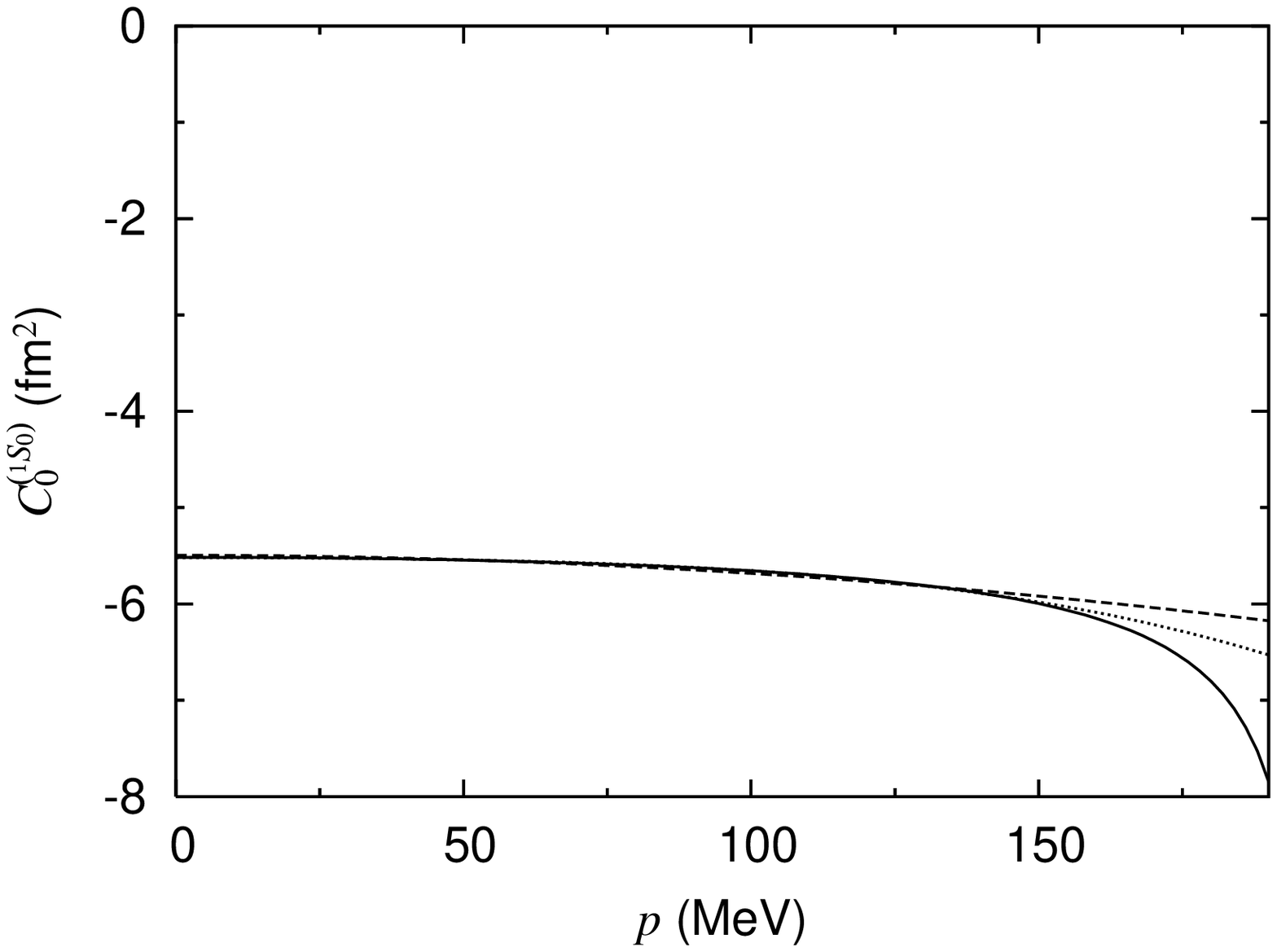}
 \caption{\label{fig_c0-dep}
The on-shell energy dependence of $C_{0}^{(\1s0)}$
contained in \vpiII\ [see \Eq{vpi_1s0}, for the definition] for
$\Lambda$ = 200 MeV.
The solid curve is obtained by fitting to $V_M$ while the dashed
 (dotted) curve is obtained by fitting to the solid one by adjusting two
 (three) parameters. (See the text for more explanation about the fitting.)
}
\end{minipage}
\hspace{5mm}
\begin{minipage}[t]{65mm}
 \includegraphics[width=65mm]{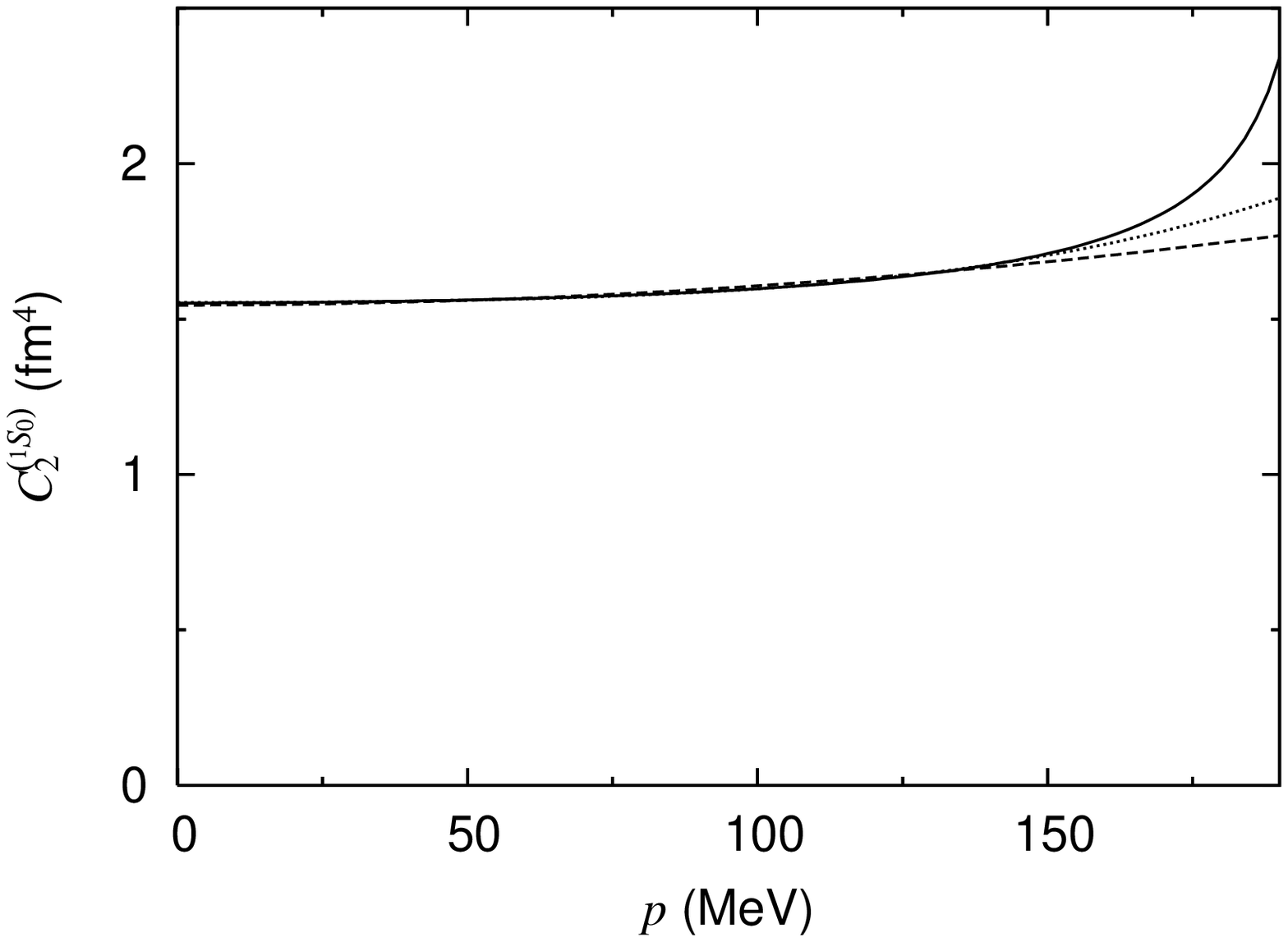}
 \caption{\label{fig_c2-dep}
The on-shell energy dependence of $C_{2}^{(\1s0)}$
contained in \vpiII\ for $\Lambda$ = 200 MeV.
The other features are the same as those in Fig.~\ref{fig_c0-dep}.
 }
\end{minipage}
\end{figure}

Now we know that it is important to consider the on-shell energy dependence
of the couplings of the contact interactions in \veft.
If we also parameterize the on-shell energy dependence,
we have to modify the counting rule.
A parameterization for
a pionless on-shell energy dependent $NN$-potential is
\begin{eqnarray}
 V(k',k;p,\Lambda)=C^\Lambda_{00}+C^\Lambda_{20}\left(k^2+k'^2\right)+C^\Lambda_{02}\,p^2+\cdots,
\end{eqnarray}
as given in Eq.~(1) of Ref.~\citen{birse}. 
A natural choice is to use a modified counting rule
which regards $p^{2n_p}k^{2n_k}$ and $p^{2m_p}k^{2m_k}$ terms 
($n_p+n_k=m_p+m_k$) as the same order.
In order for the modified counting rule to be effective,
it is necessary that the on-shell energy
dependence of the LECs can be expanded in a convergent power series in $p^2$.
In fact, such an expansion is expected to exist.
This is because the WRG equation [\Eq{rge}] indicates that
the shift of the potential due to a change of the cutoff
has an on-shell energy dependence
which can be expressed as a convergent expansion in terms of $p^2$,
if the starting potential has such a dependence.
Therefore,
if we start with a potential without on-shell energy dependence and
reduce the model space according to the WRG equation,
the obtained model-space potential would have such a tractable
energy dependence.
Indeed, we can see from Figs. \ref{fig_c0-dep} and \ref{fig_c2-dep} 
that the LECs have such on-shell energy dependence.
In the figures, we plot the on-shell energy dependence of the LECs
contained in \vpiII\ ($\Lambda$ = 200 MeV)
for $np$ $\1s0$-scattering; the definition of the LECs is given in
\Eq{vpi_1s0}.
The solid curves are obtained by fitting the LECs to $V_M$ for each $p$.
We simulate the $p^2$-dependence of the LECs using
\begin{eqnarray}
 \eqn{lec_p_dep}
 C_{2n}^{(\1s0)}(p^2) &=&  C_{2n,0}^{(\1s0)} +C_{2n,2}^{(\1s0)}p^2
  +C_{2n,4}^{(\1s0)}p^4 + \cdots\ .
\end{eqnarray}
In order to obtain the dashed (dotted) curve in the figures,
we took the  first two (three) terms in the r.h.s. of \Eq{lec_p_dep} and
fitted their parameters to the solid curve 
using a least square method
in the region $0\le p\le 150$ MeV.
We restrict the $p$-region in the simulation for the following reasons:
(1) the LECs suddenly change as functions of $p$ near
$p\sim\Lambda=$ 200 MeV, and the higher-order terms in \Eq{lec_p_dep} are
necessary for a good simulation;
(2) inclusion of the high-$p$ region in the simulation makes the
simulation worse even in the low-$p$ region.

Next we discuss the model independence of NEFT.
As mentioned in the Introduction, the model independence of NEFT
is based on the following two facts:
(1) the most general Lagrangian consistent with the assumed
symmetries (which consists of effective d.o.f.) is used;
(2) a systematic and perturbative calculation is performed following a
counting rule, and {\it naturalness}, {\it systematicness} and {\it
integrability} are necessary assumptions to organize the perturbation.
Although the validity of these assumptions had not previously been fully confirmed,
we found that they are indeed realized in the {\it proper} \veft.
Therefore, our finding supports the model independence of NEFT based on
the above-mentioned two facts.
However, we have the following question:
Although NEFT is a model-independent framework in this sense, 
is \veft\ still one of many phase-shift equivalent
potentials defined in the same model space? The answer is `yes'. 
As we have seen, the model dependence of a potential with a sufficiently reduced model space
does {\it not} arise from the description of small scale phenomena.
Instead, it arises from the choice of
the model-space reduction scheme, and this is why the answer is `yes'.
However, we have no choice of the reduction scheme
if we work within NEFT.
In order to maintain consistency with the basic ingredients of NEFT,
we must choose the WRG method.
With this renormalization scheme, there is an essentially unique model-space
potential to be simulated by \veft.

Lastly, we emphasize one point:
We should be careful that \veft\ is
defined in a model space with a relatively small cutoff.
Of course, some aspects regarding this point have been treated in
previous works.
For example, the cutoff dependence of NEFT-based predictions of
observables have been often examined.
However, there has been some confusion resulting from an insufficient
understanding of this point.

One point of confusion is seen in a study of resonance saturation.
In Ref.~\citen{resonance_sat}, the authors started with a phenomenological
one-boson-exchange potential
and directly expanded the one-boson exchange terms, excluding the OPEP,
in terms of a series of contact interactions and 
compared the couplings obtained
in this way with those obtained using the {\it conventional} procedure of \pieft.
However, this comparison makes no sense.
The potential obtained in this naive expansion cannot reproduce the
scattering observables generated by the original potential.
It should be noticed that \vph\ and \veft\
are defined in different model spaces and that therefore they cannot be
compared with each other directly.
In contrast to the above-stated manner, it is necessary
to perform the model-space reduction before the expansion.

Another point of confusion is seen in the determination of the couplings of the contact
interactions, which is relevant to the $^3S_1$-$^3D_1$ partial wave
scattering, with the use of the deuteron properties.
Some authors have used the deuteron binding energy (B.E.), the asymptotic
$S$-wave normalization ($A_S$), and the $D/S$-ratio ($\eta$) to determine the
couplings.\cite{NN-park}
However, as we discussed in the preceding section, 
$A_S$ and $\eta$ include information concerning the {\it details} of small scale
physics, and therefore they are not well-defined quantities
in the model-space NEFT framework.
We should recognize that
the use of $A_S$ and $\eta$ to determine the LECs means
that we must introduce a model into the NEFT framework.

\subsection{The {\it conventional} \veft}
\label{subsec_d2}

We now discuss the nature of
the {\it conventional} \veft\
constructed in the previous work
from our point of view.
The procedure to derive the {\it conventional} potential is
described in the Introduction. 
We describe their characteristics in the following.
They are hermitian, energy independent (except for possible higher-order
corrections) and defined in a model space with $\Lambda\ltap$1 GeV.
The unknown parameters they contain are determined such that physical
observables are accurately predicted.
Specifically, this means that
the on-shell T-matrix elements are accurately predicted.
The potential generates an orthogonal set of wave functions.
The above-stated characteristics are the same as those of a model-space
potential obtained using the UT method.
Therefore, the {\it conventional} \veft\ can be regarded as
a parametrization of a model-space potential obtained using the UT method.
Is there another interpretation?
According to the basic idea of EFT, two NEFTs defined in different
model spaces should be connected by integrating out heavier
d.o.f. 
If one wishes to reduce the model space while keeping
the {\it conventional} \veft\ equipped with
the above-mentioned characteristics, the
only known method is the UT method.
Therefore, we adopt the interpretation that
the {\it conventional} \veft\ is a parametrization of the
model-space potential due to the UT method.

With the above interpretation, we find the following problems
in the {\it conventional} \veft.
Even if this interpretation is not correct, a considerable number of the
problems would remain.
The {\it proper} \veft\ discussed in the preceding subsection, by
contrast, is free from these problems.
We now consider these problems.
In order to see the problems clearly, we often consider cases with
rather small model spaces.
First, as discussed in \S \ref{sec_rg}, we note that the UT method
is unrelated to the path integral method of integrating out
heavier d.o.f., while
an effective Lagrangian for NEFT is
considered to be obtained using the path integral method.

Secondly, as discussed in \S \ref{sec_result}, the UT method gives a
model-space potential which cannot be simulated accurately by the {\it
conventional} NEFT-based parametrization,
in particular, in the case of a small cutoff.
However, this does {\it not} mean that a {\it conventional}
on-shell energy independent \veft\
cannot predict physical observables.
For example, suppose that we construct a {\it conventional} \vpilessI,
for which
we have two contact interactions 
with zero and two derivatives, respectively.
In this case, we can find values of the LECs of the contact interactions
which give the scattering length and the effective range.
As seen in Fig.~\ref{fig_vlowk_70}, however, such a potential
(the dashed curve) cannot simulate the potential obtained with the
UT method (the solid curve),\footnote{
The solid curve in Fig.~\ref{fig_vlowk_70} is obtained from the
\vlowk\ method.
It is noted, however,  that the \vlowk\ method and the UT method yield
very similar low-momentum potentials.
}
and it is applicable to only a
very limited energy range that is far below the cutoff.
Although we have freedom in choosing the unitary transformation,
the small applicable energy range indicates that no unitary
transformation gives a potential that can be accurately simulated by the two
contact interactions.
This is an example of a case in which the {\it naturalness} of the LECs is
realized but the {\it systematicness} is not.
We should say that a low-momentum potential, like that represented by
the dashed curve in Fig.~\ref{fig_vlowk_70}, is simply a
phenomenological model meant only to reproduce the scattering length and
the effective range.
Often, in the situation that
a prediction made using the {\it conventional} NEFT begins to deviate
from the data as the energy is increased,  
it is claimed that the energy is beyond the region to which 
NEFT is applicable.
However, as we have seen, the reason that there is a small applicable energy
region is that the {\it conventional} NEFT parametrization is not suitable to
simulate the UT-generated model-space potential.

\begin{figure}
\begin{center}
 \includegraphics[width=130mm]{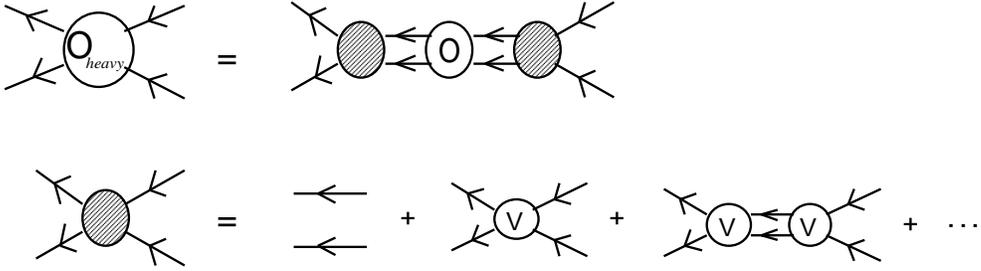}
\end{center}
\caption{\label{fig_oheavy}
A diagrammatic representation of ${\cal O}_{heavy}$
introduced in \Eq{me_eft}.
An operator ${\cal O}$ acts on the full space, while 
${\cal O}_{heavy}$ acts on the model space including only low-momentum states.
The shaded blob includes the free two-nucleon propagator plus
the ladder diagrams generated by the full-space $NN$-potential, $V$.
In all of the intermediate states involved in the diagrams,
the nucleons run over the complementary space 
of the nucleon momentum states.
}
\end{figure}

Thirdly, the UT method does not preserve the values of the off-shell T-matrix
elements. 
This leads to a (partial) breakdown of the EFT idea of the separation of
scales.
We now consider an example as an illustration.
We start with the full space and
consider a matrix element of an operator ${\cal O}$ corresponding to a
two-nucleon state $\Psi$, {\it i.e.}, $\bra{\Psi} {\cal O} \ket{\Psi}$.
First, we discuss 
the evaluation of this matrix element using
the {\it proper} NEFT.
In this method,
we begin by decomposing the matrix element
into low- and high-energy parts, as
\begin{eqnarray}
 \bra{\Psi} {\cal O} \ket{\Psi}
 = (\bra{\Psi_L} + \bra{\Psi_H}) {\cal O} (\ket{\Psi_L}
 + \ket{\Psi_H}) ,
\end{eqnarray}
with
\begin{eqnarray}
\ket{\Psi_L} = \eta \ket{\Psi}\ ,\quad \ket{\Psi_H} = \lambda \ket{\Psi} \ ,
\end{eqnarray}
where $\eta$ and $\lambda$ are the projection operators defined in
\Eqs{eta}{lambda}, respectively, and
the subscripts $L$ and $H$ indicate the low- and high-energy components,
respectively.
Then, we proceed as
\begin{eqnarray}
 \eqn{me_eft}
 \bra{\Psi} {\cal O} \ket{\Psi}
 &=& \bra{\Psi_L} {\cal O} \ket{\Psi_L} + {\rm the\ other\ terms}\\\nonumber
 &=& \bra{\Psi_L} {\cal O} \ket{\Psi_L} + 
\bra{\Psi_L} {\cal O}_{heavy} \ket{\Psi_L}\\\nonumber
 &\simeq& \bra{\Psi_L} {\cal O} \ket{\Psi_L} + 
\bra{\Psi_L} {\cal O}_{cnt} \ket{\Psi_L}\ ,
\end{eqnarray}
where ${\cal O}_{heavy}$ is given by one insertion of ${\cal O}$ with
all orders of rescattering terms due to the $NN$-potential;
all intermediate states involved are the high-momentum states contained
in the projection operator $\lambda$ (see Fig.~\ref{fig_oheavy} for a
diagrammatic representation of ${\cal O}_{heavy}$).
Further, ${\cal O}_{cnt}$ is a series of contact operators and their
couplings are generally dependent on the on-shell energies of the
initial and final states. They are determined such that the best
simulation of ${\cal O}_{heavy}$ is realized.
The step from the first to the second line in \Eq{me_eft}
indeed represents the model-space reduction according to the WRG equation.
The step from the second to the third line is made by carrying out an
expansion of ${\cal O}_{heavy}$ in terms of the contact operators.
The third line is what we consider in NEFT.
The low-energy physics, namely, that described by
$\bra{\Psi_L} {\cal O}\ket{\Psi_L}$ is
treated identically to the original theory.
The separation of scales is clearly realized, which we expect
for EFT.

Next, we consider the case for the UT method. Here we have
\begin{eqnarray}
 \eqn{me_ut}
 \bra{\Psi} {\cal O} \ket{\Psi}
 &=& \bra{\Psi} U U^\dagger {\cal O} U U^\dagger \ket{\Psi}\\\nonumber
 &=& \bra{\Psi^\prime} ( {\cal O} + V^\dagger{\cal O} + {\cal O}V 
 + V^\dagger{\cal O}V ) \ket{\Psi^\prime}\\\nonumber
 &\rightarrow& \bra{\Psi^\prime} {\cal O} \ket{\Psi^\prime}
 + \bra{\Psi^\prime} {\cal O}_{cnt}^\prime \ket{\Psi^\prime}\ ,
\end{eqnarray}
where $U$ is the unitary operator
introduced in \Eq{transf} to reduce the model space, and we have used
the definitions
$\Psi^\prime\equiv U^\dagger\Psi\ne \eta \Psi$ and $V\equiv U-1$.
The step from the first to the second line represents
the model-space reduction using the UT method.
The third line is what one evaluates in the {\it conventional} NEFT.
As we see, 
the low-energy physics is {\it not} treated in
the same way as in the original theory; the high-energy physics enters
through $\bra{\Psi^\prime} {\cal O} \ket{\Psi^\prime}$, while
the low-energy physics is mixed in 
$\bra{\Psi^\prime} {\cal O}_{cnt}^\prime \ket{\Psi^\prime}$.
In fact, the UT method causes
mixing of scales rather than separation of scales.
This is why, as stated above, a clear separation of scales cannot be
accomplished with the UT method.
Actually, the situation is even more serious.
In the case that ${\cal O}$ is a Hamiltonian, as we have seen in our result, 
the step from the second to the third line in \Eq{me_ut}
(the expansion in terms of contact interactions) cannot be carried out
satisfactorily. 
For the other operators, such as electromagnetic currents, the same
difficulty may exist.

The following arguments hold regardless of whether the {\it
conventional} NEFT is related to the UT method.
In the {\it conventional} NEFT, one starts from a model space.
Thus, what one does in the {\it conventional} NEFT is not to follow all
the steps taken in \Eq{me_ut} but to start from the third line.
In this case, $\Psi^\prime$ is generated by a potential whose role is
only to
reproduce on-shell quantities for a rather limited energy region,
as discussed in the previous paragraph.
This means that the off-shell behavior of $\Psi^\prime$ is completely
out of our control. 
What one can do in the {\it conventional} NEFT is to determine
the couplings involved with ${\cal O}^\prime_{cnt}$
at a given kinematical point by fitting to data and hope
that this uncontrollable behavior is completely canceled by
${\cal O}^\prime_{cnt}$.
However,
all the {\it ad hoc} steps taken in the {\it conventional} NEFT,
as discussed above,
are expected to shrink the kinematical region in which the theory is valid.
The {\it proper} NEFT should have much broader applicability.

Despite the weakness of
the {\it conventional} NEFT described above, it has had much
success.
We now consider the reason it seems to have been successful.
A likely reason is that most of the previous works have been concerned
with \eftpi, and a relatively large cutoff value was used.\footnote{
There is a well-known success of \piless\ with
KSW-counting,\cite{pds} in which dimensional regularization plus the
so-called power divergence subtraction is used.
}
As we have seen in Fig.~\ref{fig_vlowk_400},\footnote{
It is noted that
the result for the UT method is essentially
the same as that for the \vlowk\ method.
}
the behavior of the UT-generated
model-space potential around the cutoff becomes much smoother for a larger
cutoff. Therefore, it is practically possible in this case to realize a
more accurate simulation of the model-space potential in
terms of one- (and two-) pion-exchange plus contact interactions.
Furthermore, it is noted that
$V_M$'s obtained from the WRG and the \vlowk\ methods are very similar
for a larger cutoff, as we found in Fig.~\ref{fig_vlowk_400}.
Therefore, the {\it conventional} \vpi\ with a large cutoff
can be interpreted as a parameterization of the WRG-based $V_M$,
which also supports the practical use of the {\it conventional} procedure.
On the other hand,
it is {\it not} expected that \vpiless\
constructed in the {\it conventional} manner is effective,
because the pionless potential
should be defined in a rather small model space.
However,
the usefulness of EFT should not depend on the value of the
cutoff if appropriate d.o.f. for the cutoff are considered explicitly.
Even though the {\it conventional} \pieft\ with a large cutoff
is practically useful, it is noted that this fact
does not completely justify the {\it conventional} NEFT.
%

Lastly, we make one more comment on the success of the {\it
conventional} \veft.
Although the scattering observables have been thoroughly studied
with the use of \veft,
electroweak processes have not,\footnote{
Although there have been many NEFT-based studies of electroweak processes
in few-nucleon systems,
a \veft\ constructed in the {\it conventional} manner has not been used
directly in those studies.
}
and the off-shell behavior of the wave functions
generated by \veft\ have not yet been thoroughly examined.
As we have discussed, even if some potentials can accurately predict the
scattering
phase shifts, they do not always control the off-shell behavior.
We should consider the success of the {\it conventional} NEFT
while keeping this point in mind.

\subsection{What have we learned?}
\label{subsec_d3}

In this work, we have explicitly demonstrated that the RG idea is valid
in application to the nuclear force.
The validity of the RG idea 
is a condition for NEFT to be effective.
The RG idea is that
as we reduce the model space sufficiently, or
equivalently, as we view a system in a sufficiently coarse-grained
manner, we find the system to
be controlled by a dynamics
that are independent of the
{\it details} of the small scale physics.
Although we may need a confirmation,
it is likely that the RG idea is valid in application to 
two-nucleon systems as a whole, {\it i.e.}, in application to
the nuclear current operator as well as the nuclear force.
In the following, we consider what we can learn about 
a description of a two-nucleon system based on 
a traditional, phenomenological model approach,
and about a description based on 
NEFT from a point of view rooted in the RG idea in
a two-nucleon system.

A model used in the traditional approach
contains well-known large scale physics and 
phenomenological small scale physics.
There are many such models which differ from
each other with regard to small scale physics.
Those models are able to describe a given low-energy reaction with
similar precision.
The RG approach tells us that these models are equivalent at low energies;
all of these models are, after an appropriate model-space reduction,
transformed into essentially a single low-energy theory,
which does not have the dependence on the modeling of the small scale
physics. This low-energy theory can be interpreted as NEFT.
Therefore, the RG procedure guarantees that the models and NEFT give the same
model-independent result in a low-energy
region in which the {\it details} of small scale physics involved in the
models are unimportant. 
We do not have to concern ourselves with the model dependence
of the predictions obtained with the models, 
as long as we are concerned with that energy region.
This point of view based on the RG idea
may be interpreted as a foundation of the traditional
model approach.

As we have seen, the RG idea makes the relation between NEFT and 
the model approach clear.
From this viewpoint, we give a supplementary explanation for the {\it
conventional} argument regarding the interplay between NEFT and the
model approach.
Typically, models are constructed in a model space that is
much larger than those for NEFT.
Therefore, these models are, in principle, applicable to energy regions
outside the region to which NEFT is applicable.
It has sometimes been claimed that 
it is a reasonable strategy to apply a model to a higher energy region if
its reliability had been tested in the low-energy region
by comparing its predictions with NEFT-based ones.
As we go into higher energy regions, however,
the dependence on the modeling of small scale physics gradually
enters into model predictions.
Because the {\it details} of the small scale physics involved in the models
has never been tested by the comparison, this type of extrapolation 
is expected to be rather limited.
For a better extension of the model to higher energy regions, it is
important to test the small scale physics part by using experimental
data from that energy region.

With the understanding of the relation between NEFT and the traditional
model approach,
we now discuss what is, and is not, newly gained by working with NEFT.
One gain is that NEFT enables us to perform a perturbative calculation.
As a consequence, we can improve the accuracy of a calculation systematically
and we can estimate the theoretical uncertainty.
It is noted, however, that there are cases in which the perturbation
expansion does not converge sufficiently rapidly. 
This is because the contribution of the leading terms is suppressed by a
certain symmetry.
An example is the $pp\rightarrow pp\pi^0$ reaction.\cite{pp_pppi0}
Another gain is that we can work with interactions free from the
model dependence that results from treating the details of the
small scale physics.
This does not mean, however, that only NEFT gives a model-independent
result. 
In the energy region in which NEFT is useful, NEFT and 
a reasonable phenomenological model are equivalent
in the sense that they are connected
through the RG, as we have seen.
Actually, we should recognize that
a new phenomenological prediction is rarely gained by working with NEFT.
We can, in many cases, construct a reasonable model which is
equivalent to NEFT at low energies;
this is what nuclear physicists have done.
Even so, there is one more gain, namely, that NEFT enables us to
construct a nuclear
force and nuclear electroweak current operator much more efficiently
with much simpler parametrization than 
in the case that we construct a model.
This efficiency is due to the fact that we know which interactions are to
be considered at a given order of the NEFT perturbation.
In this context, one may say that the chiral symmetry plays an important
role in \eftpi\ because the chiral symmetry determines
how the pion interacts with the other particles, and
the use of interactions satisfying 
the chiral symmetry is important in setting up the counting
rule.\footnote{
It may not be warranted to make
this typical argument, from the result obtained in this work,
about the role played by the chiral symmetry in NEFT.
Our result implies that any framework
({\it i.e.}, a Lagrangian and a perturbation scheme) is useful
if it gives a good parameterization of a model-space $NN$-potential
which is free from model dependence coming from the description of small
scale physics;
NEFT is one of them.
In order to assess the role of the chiral symmetry, 
it would be helpful to examine an $NN$-potential with $\Lambda\sim$ 400 MeV
in which the {\it details} of multi-pion-exchange potentials are
expected to play
an important role. In this case, one has to explicitly include
the multi-pion-exchange and therefore needs a rule to perturbatively
include them;
the use of the chiral Lagrangian may be important for this purpose.
If the OPEP is the only important pion-induced mechanism, as in
our result, we do not have to rely on the chiral Lagrangian, because
many $NN$-models include the OPEP without using the chiral Lagrangian.
}
In constructing a model, on the other hand, one has to find an
appropriate combination
of interactions by trial and error, except for the well-known long-range
mechanism.
The simple parameterization of interactions in EFT results from the fact
that we ignore {\it details} of small scale physics.

\section{Summary and conclusion}\label{sec_conclusion}

We have demonstrated the relation between \veft\ and \vph\ in this work.
For this purpose, we studied how to reduce the model space.
One method considered as a possibility for this reduction is
the WRG equation, which is equivalent to
the path integral method of integrating out heavier d.o.f. 
We found that the WRG equation generates a potential
which can be accurately simulated by \veft;
the basic assumptions of NEFT, namely, {\it naturalness},
{\it systematicness} and {\it integrability} are realized in
\veft.
Thus, we conclude that the WRG method is the appropriate model-space
reduction in NEFT.
Simultaneously, we conclude that \vph\ and \veft\
(and different \veft's acting on different model spaces)
are connected through the WRG.
It was shown that \veft\ is free from model dependence,
{\it i.e.}, dependence on the description used for the small scale
physics.
The use of simple contact interactions in NEFT can be naturally
understood from the RG point of view.

We also examined two other model-space reduction methods, the \vlowk\
and the UT methods.
However, we found that they are not consistent with the path integral
method and generate potentials that are not consistent with the
basic NEFT assumptions.
We conclude that they are not appropriate methods of model-space
reduction in NEFT.

After finding the relation between \veft\ and \vph,
we discussed the \veft\ properties which 
have not yet been fully recognized.
Because \veft\ is obtained by integrating out heavy d.o.f. following the
RG procedure, the \veft\ should have properties consistent with the
model-space reduction.
For example, \veft\ is on-shell energy dependent; we
proposed a method for modifying the {\it conventional} counting rule.
We emphasized the importance of being careful that \veft\ be
defined in a model space whose size is sufficiently smaller than those
used for \vph. 
We emphasized this point because there has been confusion that has
resulted from overlooking this point.

From the point of view developed in this work,
we discussed the nature of the {\it conventional} \veft.
We discussed the fact that the {\it conventional} \veft\ can be
interpreted as a
parametrization of a potential whose model space has been reduced from
some theory with the use of the UT method.
Whether or not this interpretation is correct,
the {\it conventional} \veft\ has problems from which the {\it
proper} \veft\ is free.
In particular, one serious problem is that the mixing of scales rather
than the
separation of scales is involved in the {\it conventional} \veft.
This implies that
a model is introduced into the theory to a certain extent.
These problems appear clearly when the value of the cutoff is small.
We conclude that the {\it conventional} \veft\ is not fully
consistent with the basic idea of NEFT,
but it is practically useful in the case of a large model space and
has had been successful phenomenologically.

Finally, we discussed what we can learn about NEFT and the traditional
model approach from the RG idea.
We conclude that the model approach is
equivalent at low energies to NEFT in describing low-energy two-nucleon
systems. Therefore,
we do not have to be concerned with the model dependence of predictions
made by a reasonable phenomenological model,
as long as we are concerned with energy regions in which {\it details} of
small scale physics is unimportant.
We also conclude that we can gain the following by working with NEFT:
a perturbative calculational procedure which enables
us to systematically add higher-order corrections and to estimate
theoretical uncertainty;
a model-independent framework which, however, does not mean that only NEFT
gives model-independent predictions;
an efficient and simple construction of nuclear systems, for which chiral
symmetry plays an important role in the case that the pion is dynamical.

\section*{Acknowledgements}
I would like to express my sincere thanks to Prof. H. Toki for
his critical reading of this paper and for valuable discussions on this work.
I am very grateful to Prof. T. Sato for invaluable discussions.
I would like to express my deep gratitude to Prof. K. Kubodera
for his great interest in this work, his continuous encouragement
and his letting me work at the University of South Carolina.
The initial stage of this work has been done there.
My deep thanks are also due to Prof. T. Kunihiro for useful
discussions, continuous encouragement and his giving me an
opportunity to work at the Yukawa Institute for Theoretical Physics
(YITP) at Kyoto University. Considerable parts of this work was
done during my stay at the YITP.
Finally, I give my great thanks to Prof. R. Seki,
Dr. S. Ando and Dr. H. Kajiura for very informative discussions.
This work is supported in part by the United States National Science
Foundation, Grant No.  PHY-0140214.

\appendix
\section{Derivation of the Wilsonian Renormalization Group Equation}
\label{app_rg}

In this appendix, we derive the WRG equation [\Eq{rge}] by explicitly
integrating out the nucleon high-momentum states in a path integral.
We restrict ourselves to the two-nucleon center of mass (CM) system.
Therefore, we consider only a two-body force; that is,
we do not consider the inclusion of intrinsic
multi- ({\it i.e.}, more than two) body forces nor their generation
due to integrating out heavy d.o.f.
For simplicity, we consider the case in which there is no mixing of
different partial waves.
Extension to the coupled-channel case is straightforward.
We start with a path integral $Z$ written only in terms of the nucleon
field $N$:
\begin{eqnarray}
 \eqn{z_ln}
 Z &=& \int {\cal D}N{\cal D}N^\dagger e^{i\int  d^4\!x\ {\cal L}_N}\ ,
\end{eqnarray}
with
\begin{eqnarray}
 \eqn{lagn}
{\cal L}_N &=& {\cal L}_o
- \sum_{\alpha,\lambda} D^{(\alpha)\dagger}_\lambda\ W^{(\alpha)}\ 
D^{(\alpha)}_\lambda\ ,\\
{\cal L}_o &=& N^\dagger\left(i\partial_0 + {\bm{\nabla}^2\over
    2M}\right)N \ , \\
 \eqn{lagn2}
D^{(\alpha)}_\lambda &=& N^TP_\lambda^{(\alpha)}N\ ,
\end{eqnarray}
where $\alpha$ specifies a partial wave of an $NN$-state and 
$P_\lambda^{(\alpha)}$ denotes the projection operator onto a partial
wave $\alpha$.
[Explicit expressions for $P_\lambda^{(\alpha)}$ are given in Eq.~(A4) of
Ref.~\citen{projection}.] The suffix $\lambda$ indicates the direction of
the spatial and isospin-spatial polarization of the two-nucleon system.
The quantity $W^{(\alpha)}$ is the coupling of the two-body
$NN$-interaction for a partial wave $\alpha$.
$W^{(\alpha)}$ contains derivatives and therefore
depends on the magnitudes of the momenta carried by
the free incoming nucleons ($k$) and outgoing nucleons ($k'$) at vertex;
it can also depend on the on-shell momentum ($p$) of the two-nucleon
system and the momentum cutoff ($\Lambda$).
We can thus write
$W^{(\alpha)}=W^{(\alpha)}(k',k;p,\Lambda)$.
The arguments of $W^{(\alpha)}$ are suppressed until \Eq{rge2}
for simplicity.
The quantity $V^{(\alpha)}$ defined by 
$V^{(\alpha)}\equiv W^{(\alpha)}/2$
is interpreted as an $NN$-potential.

We separate the nucleon field into high-frequency modes $N_H$ and
low-frequency modes $N_L$.
We then expand the high-frequency modes of the interaction Lagrangian to obtain
\begin{eqnarray}
 \eqn{z_ln2}
 Z &=&
 \int {\cal D}N_L{\cal D}N_L^\dagger\ 
 e^{i\int  d^4\!x\ {\cal L}_{N}^{(L)}}
\int {\cal D}N_H{\cal D}N_H^\dagger\ 
 e^{i\int  d^4\!x\ {\cal L}_o^{(H)}}\Bigg\{1 \\\nonumber
&+&{(-i)^2\over 2!}\sum_{\alpha\beta,\lambda\rho} \int d^4x_1 d^4x_2
D^{(\alpha)\dagger}_{\!L\, \lambda}(x_1)\ W^{(\alpha)}\ D^{(\alpha)}_{\!H\, \lambda}(x_1)
D^{(\beta)\dagger}_{\!H\, \rho}(x_2)\ W^{(\beta)}\ D^{(\beta)}_{\!L\, \rho}(x_2)
 + \cdots \Bigg\} \ ,
\end{eqnarray}
where ${\cal L}_{N}^{(L)}$ includes only the nucleon fields with
low-frequency modes.
Integration of the first term in
\Eq{z_ln2} over $N_H$ leads to an unimportant overall factor of $Z$.
We explicitly perform the integration of the second term in the
following. This integration leads to
the one-loop correction to the two-body interaction.
The ellipsis includes all possible diagrams of products of connected
multi-loop ladder pieces.
The self-energy pieces are not considered, because we should maintain
consistency with the NEFT calculational method.
In the NEFT calculation, one solves the Lippmann-Schwinger equation,
which means that only the ladders of the $NN$-potential are resummed.

Integrating the second term above yields
\begin{eqnarray}
 \eqn{z_ln3}
&&{\int {\cal D}N_H{\cal D}N_H^\dagger\ 
 e^{i\int  d^4\!x\ {\cal L}_o^{(H)}}
D^{(\alpha)}_{\!H\, \lambda}(x_1)D^{(\beta)\dagger}_{\!H\, \rho}(x_2)
\bigg/ \int {\cal D}N_H{\cal D}N_H^\dagger\ 
 e^{i\int  d^4\!x\ {\cal L}_o^{(H)}}}\\\nonumber
&=&\int_H {d^4q_1\over (2\pi)^4}{d^4q_2\over (2\pi)^4}\left(
{ie^{iq_1\cdot(x_2-x_1)}\over q_1^0-\bm{q}_1^2/2M+i\epsilon}
{ie^{iq_2\cdot(x_2-x_1)}\over q_2^0-\bm{q}_2^2/2M+i\epsilon}
\right)
\left(2 {\rm Tr} \left(P_{\lambda}^{(\alpha)}P_{\rho}^{(\beta)\dagger}\right)
\right)\ ,
\end{eqnarray}
where $\int_H$ indicates that $\bm{q}_i\ (i=1,2)$,
the nucleon momentum in the intermediate state, runs over the region for
the high-frequency modes.
In \Eq{z_ln3}, we have used the relation
$P_{\lambda}^{(\alpha)T}=-P_{\lambda}^{(\alpha)}$.
Now, \Eq{z_ln2} becomes, up to an unimportant overall factor,
\begin{eqnarray}
 \eqn{z_ln4}
 Z &=&
 \int {\cal D}N_L{\cal D}N_L^\dagger\ 
 e^{i\int  d^4\!x\ {\cal L}_{N}^{(L)}}\Bigg\{
1+{1\over 2}\sum_{\alpha\beta,\lambda\rho} 
\int_H {d^4q_1\over (2\pi)^4}{d^4q_2\over (2\pi)^4}
\int d^4x_1 d^4x_2\\\nonumber
&\times& 
D^{(\alpha)\dagger}_{\!L\, \lambda}(x_1)
W^{(\alpha)}\ 
{2 {\rm Tr}
\left(P_{\lambda}^{(\alpha)}P_{\rho}^{(\beta)\dagger}\right)\,
e^{i(q_1+q_2)\cdot(x_2-x_1)}\over \left(q_1^0-\bm{q}_1^2/2M+i\epsilon\right)
\left(q_2^0-\bm{q}_2^2/2M+i\epsilon\right)}\
W^{(\beta)}
D^{(\beta)}_{\!L\, \rho}(x_2)
+\cdots
\Bigg\}\\\nonumber
%
&=& \int {\cal D}N_L{\cal D}N_L^\dagger\ 
 e^{i\int  d^4\!x\ {\cal L}_{N}^{(L)}}\Bigg\{
1+{1\over 2}(2\pi)^4\delta^{(4)}(P-P^\prime)\sum_{\alpha\beta,\lambda\rho}
D^{(\alpha)\dagger}_{\!L\, \lambda}(0)
W^{(\alpha)}\ \\\nonumber
&\times&
\int_H {d^4q_1\over (2\pi)^4} 
{2 {\rm Tr}
\left(P_{\lambda}^{(\alpha)}P_{\rho}^{(\beta)\dagger}\right)
\over 
\left(q_1^0-\bm{q}_1^2/2M+i\epsilon\right)
\left(P^0-q_1^0-(\bm{P}-\bm{q}_1)^2/2M+i\epsilon\right)}\,
W^{(\beta)}
D^{(\beta)}_{\!L\, \rho}(0)
+\cdots
\Bigg\}\\\nonumber
&=& \int {\cal D}N_L{\cal D}N_L^\dagger\ 
 e^{i\int  d^4\!x\ {\cal L}_{N}^{(L)}}\Bigg\{
1-{i\over 2}(2\pi)^4\delta^{(4)}(P-P^\prime)\\\nonumber
&\times&
\sum_{\alpha\beta,\lambda\rho}
D^{(\alpha)\dagger}_{\!L\, \lambda}(0)
W^{(\alpha)}\ 
\int_H {d^3q_1\over (2\pi)^3} 
{2 {\rm Tr}
\left(P_{\lambda}^{(\alpha)}P_{\rho}^{(\beta)\dagger}\right)
\over P^0-\bm{q}_1^2/M+i\epsilon}
W^{(\beta)}
D^{(\beta)}_{\!L\, \rho}(0)
+\cdots
\Bigg\}\ , 
\end{eqnarray}
where we have used the fact that we are working in the CM system.
The energy and momentum of the initial (final) two-nucleon system is
denoted $P=(P^0,\bm{P})$ ($P'$).
We consider the case in which an infinitesimally small momentum
shell is integrated out; that is, $\bm{q}_1$ in \Eq{z_ln4} runs over a region
$\Lambda-\delta\Lambda\le|\bm{q}_1|\le\Lambda$.
In this case, use of the relation
$\int d\Omega_{\bm{q}_1} 2 {\rm Tr}
\left(P_{\lambda}^{(\alpha)}P_{\rho}^{(\beta)\dagger}\right)
=4\pi\,\delta_{\alpha\beta}\,\delta_{\lambda\rho}$ leads to
\begin{eqnarray}
 \eqn{z_ln5}
 Z &=&
\int {\cal D}N_L{\cal D}N_L^\dagger\ 
 e^{i\int  d^4\!x\ {\cal L}_{N}^{(L)}}\Bigg\{
1-{i\over 2}(2\pi)^4\delta^{(4)}(P-P^\prime)\\\nonumber
&\times&
\sum_{\alpha,\lambda}
D^{(\alpha)\dagger}_{\!L\, \lambda}(0)\left(
W^{(\alpha)}\ 
{1\over 2\pi^2} 
{\Lambda^2\delta\Lambda\over P^0-\Lambda^2/M}
W^{(\alpha)}+{\cal O}\left((\delta\Lambda)^2\right)
\right)
D^{(\alpha)}_{\!L\, \lambda}(0)
\Bigg\}\\\nonumber
&=&
\int {\cal D}N_L{\cal D}N_L^\dagger\ 
 e^{i\int  d^4\!x\ {\cal L}_{N}^{(L)}}
\left\{1-i \sum_{\alpha,\lambda} \int d^4x 
D^{(\alpha)\dagger}_{\!L\, \lambda}(x)\left(
\delta W^{(\alpha)}
+{\cal O}\left((\delta\Lambda)^2\right)
\right)
D^{(\alpha)}_{\!L\, \lambda}(x)
\right\}\ ,
\end{eqnarray}
with
\begin{eqnarray}
\delta W^{(\alpha)}
&=&
{1\over 4\pi^2} 
W^{(\alpha)}
{\Lambda^2\delta\Lambda\over P^0-\Lambda^2/M}
W^{(\alpha)}\ .
\end{eqnarray}
By resumming the terms in the curly brackets in \Eq{z_ln5} as an
exponential, we obtain the path integral
\begin{eqnarray}
 \eqn{z_ln6}
 Z &=&
\int {\cal D}N_L{\cal D}N_L^\dagger\ 
 e^{i\int  d^4\!x\ {\cal L}_{N, eff}^{(L)}}\ ,
\end{eqnarray}
with
\begin{eqnarray}
{\cal L}_{N, eff}^{(L)} &=& {\cal L}_{N}^{(L)}
-i \sum_{\alpha,\lambda}
D^{(\alpha)\dagger}_{\!L\, \lambda}(x)
\left(
\delta W^{(\alpha)}+{\cal O}\left((\delta\Lambda)^2\right)
\right)
D^{(\alpha)}_{\!L\, \lambda}(x)
\\\nonumber
&=& {\cal L}_{o}^{(L)}
-i \sum_{\alpha,\lambda}
D^{(\alpha)\dagger}_{\!L\, \lambda}(x)
\left(W^{(\alpha)}+\delta W^{(\alpha)}
+{\cal O}\left((\delta\Lambda)^2\right)\right)
D^{(\alpha)}_{\!L\, \lambda}(x)\ .
\end{eqnarray}
Therefore, for an infinitesimally small reduction of the momentum
cutoff, we have the following renormalization group equation for
$W^{(\alpha)}$:
\begin{eqnarray}
 \eqn{rge2}
-{\partial W^{(\alpha)}(k',k;p,\Lambda)
\over \partial\Lambda}
&=&
{M\over 4\pi^2} 
W^{(\alpha)}(k',\Lambda;p,\Lambda)
{\Lambda^2\over p^2-\Lambda^2}
W^{(\alpha)}(\Lambda,k;p,\Lambda)\ .
\end{eqnarray}
Here, the minus sign on the l.h.s. indicates that the r.h.s. is the
 shift of $W^{(\alpha)}$ due to an infinitesimal {\it decrease} of
 $\Lambda$. We have also used $P^0=p^2/M$.
Thus we obtain the WRG equation in terms of 
$V^{(\alpha)}$ introduced above:
\begin{eqnarray}
{\partial V^{(\alpha)}(k',k;p,\Lambda)
\over \partial\Lambda}
&=&
{M\over 2\pi^2} 
V^{(\alpha)}(k',\Lambda;p,\Lambda)
{\Lambda^2\over \Lambda^2-p^2}
V^{(\alpha)}(\Lambda,k;p,\Lambda)\ .
\end{eqnarray}

\section{$NN$-Potential}
\label{appendix_NN}

We now present expressions for \vpi\ used in this work
in the partial-wave basis.
These are obtained from \Eq{vpi} by changing the basis.
The expressions presented here are substituted directly into the
Lippmann-Schwinger equation, \Eq{lippmann}.
The result for the $\1s0$ channel, $V^{(\1s0)}\equiv\bra{\1s0 }V\ket{\1s0}$,
is
\begin{eqnarray}
 \eqn{vpi_1s0}
V^{(\1s0)}(k',k;p,\Lambda)
&=&  {g_A^2\over 4f_\pi^2}\left\{-{L_1(\gamma)\over 2}
+{k^2+k'^2\over 4kk'}L_0(\gamma)\right\}\\\nonumber
&+&  C_0^{(\1s0)} + C_2^{(\1s0)} (k^2+k'^2)
+ C_4^{(\1s0)} \left\{(k^2+k'^2)^2 + (2kk')^2/3\right\}\ ,
\end{eqnarray}
with
\begin{eqnarray}
 L_J(\gamma)&\equiv& \int^1_{-1}dt {P_J(t)\over \gamma-t}\ ,
\end{eqnarray}
and $\gamma\equiv (k^2+k'^2+m_\pi^2)/2kk'$.
Here, $P_J(t)$ is a Legendre function of order $J$.
The other symbols are the same as those used in the text.
We use the average value for the pion
mass, $m_\pi=(m_{\pi^+}+m_{\pi^-}+m_{\pi^0})/3$, and we adopt
the phenomenological value
$g_A^2/4f_\pi^2 = 0.075\cdot 4\pi/m_\pi^2$.
The LECs of the contact interactions depend on $\Lambda$ and $p$.
The values of the LECs for some sets of $\Lambda$ and $p$
are given in Table \ref{tab_lecs}.
The result for
$V^{(\ts1)}(k',k;p,\Lambda)\equiv\bra{k^\prime,\td1 }V\ket{k,\ts1}$
 is obtained by simply replacing 
$C_i^{(\1s0)}$ in \Eq{vpi_1s0}
with $C_i^{(\ts1)}$ ($i=0,2,4$).
The result for $V^{({}^3D-S_1)}\equiv\bra{\td1 }V\ket{\ts1}$ is
\begin{eqnarray}
 \eqn{vpi_eps}
 V^{({}^3D-S_1)}(k',k;p,\Lambda)
&=& {g_A^2\over 4f_\pi^2}\sqrt{2}\left\{
-L_1(\gamma)+{k'\over 2k}L_0(\gamma)+{k\over 2k'}L_2(\gamma)\right\}\\\nonumber
&+&  D_2^{(\epsilon_1)} k'^2 + D_4^{(\epsilon_1)} k'^2 
\left({7\over 3}k^2 + k'^2\right)\ ,
\end{eqnarray}
and interchanging $k$ and $k'$ on the r.h.s. gives the
expression for $\bra{k^\prime,\ts1 }V\ket{k,\td1}$.
Finally, the result for $V^{(\td1)}\equiv\bra{\td1 }V\ket{\td1}$ is
\begin{eqnarray}
 V^{(\td1)}(k',k;p,\Lambda) &=&
{g_A^2\over 4f_\pi^2}\left\{
{L_1(\gamma)\over 2}-{k^2+k'^2\over 4kk'}L_2(\gamma)\right\}
+{8\over 15}C_4^{(\td1)}k^2k'^2\ .
\end{eqnarray}

\end{document}